\documentstyle[11pt,aps,tighten,preprint]{revtex}

\input epsf.sty
\input epsfig.sty

\begin{document}

\preprint{\vbox{Submitted to Physical Review D
 \hfill  UA/NPPS-8-98 \\
                \null\hfill\ FZJ-IKP(Th)-1998-14 \\
                \null\hfill\ IASA 98-1 \\ }}

\title {\bf  Chiral symmetry and the Delta-Nucleon
             transition form factors  }

\author{ G.~C.~Gellas$^1$, T.~R.~Hemmert$^{2}$, C.~N.~Ktorides$^1$,
        G.~I.~Poulis$^{1,3}$
       }
\address {
         $^1$  University of Athens,  Department of Physics\\
               Nuclear and Particle Physics Section\\
               Panepistimiopolis\\
               GR-15771 Athens,
               Greece\thanks{e-mail: ggellas@atlas.cc.uoa.gr}\\
        $^2$ Forschungszentrum J{\" u}lich, IKP (Th.) \\
             D-52425 J{\" u}lich,
             Germany\thanks{email: th.hemmert@fz-juelich.de}
             \\
        $^3$ Institute of Accelerating Systems and Applications\\
             Panepistimiopolis, Athens,
             Greece\thanks{e-mail: gpoulis@rtm.iasa.uoa.gr}\\[2ex]
         }
\medskip
\date{\today}

\maketitle

\begin{abstract}{\em }

The three complex form factors entering
the $\Delta\rightarrow N\gamma^\ast$ vertex
are calculated to ${\cal O}(\varepsilon^3)$ in the framework of a
chiral effective theory with explicit $\Delta$(1232) degrees of freedom
included.
It is shown that the low $q^2$
behavior of the form factors is governed by $\pi N$, $\pi\Delta$ loop
effects.
Predictions are given for the $q^2$-dependence of the three
transition-multipoles
$M1(q^2),\,E2(q^2),\,C2(q^2)$. Furthermore, the role
of presently unknown low energy constants that affect the values of the
multipole ratios $EMR(q^2)$ and $CMR(q^2)$ is elucidated.

\end{abstract}

\draft

\maketitle
\setlength{\topmargin}{0cm}
\evensidemargin 0mm
\oddsidemargin 0mm
\parskip 3mm plus 2mm minus 2mm
\medskip
\def\thru#1{\mathrel{\mathop{#1\!\!\!/}}}
\def\md{M_\Delta}
\def\D{\Delta}
\def\dm{M\!+\!\Delta}
\def\M{M}
\def\Mt{M^2_t}
\def\W{\Omega}
\def\aco{\arccos{\left(-{\W\over M_t}\right)}}
\def\loga{\ln{\left(-{\W\over M_t} + \sqrt{{\W^2\over M_t^2}-1}\right)}}
\def\smw{\sqrt{M_t^2-\W^2}}
\def\swm{\sqrt{\W^2-M_t^2}}
\def\naco{\arccos{\left(-{\W\over m}\right)}}
\def\nloga{\ln{\left(-{\W\over m} + \sqrt{{\W^2\over m^2}-1}\right)}}
\def\eq(#1){Eq.~(\ref{#1})}
\def\m{\mu}
\def\g5{\gamma_5}
\def\a{\alpha}
\def\n{\nu}
\def\e{\epsilon}
\def\ve{\varepsilon}
\def\dt{\cdot}
\def\nsmw{\sqrt{m^2-\W^2}}
\def\nswm{\sqrt{\W^2-m^2}}
\def\pmb#1{\setbox0=\hbox{$#1$}%
\kern-.025em\copy0\kern-\wd0
\kern.05em\copy0\kern-\wd0
\kern-.025em\raise.0433em\box0 }
\def\b{\pmb}

\section{Introduction}

The electromagnetic transition of the $\Delta$(1232) resonance to the
nucleon $\Delta\rightarrow N \gamma^\ast$
is of particular interest~\cite{exp,th,disp,currexp}
as far as our understanding of the structure of
the latter is concerned. Historically ({\it e.g.} \cite{NRQM}),
this reaction raised a lot of interest
because it allowed one to probe the issue
of whether the nucleon or its first resonance
is  ``deformed''---the reason being that apart from the dominant
magnetic dipole ($M1$) transition electromagnetic selection rules also allow
an electric ($E2$) and  a Coulomb ($C2$) quadruple transition, which vanish
in simple models of the nucleon with spherical symmetry.
Accordingly, the amount of deformation can be quantified
by the multipole ratios EMR$(q^2)=E2(q^2)/M1(q^2)$ and
CMR$(q^2)=C2(q^2)/M1(q^2)$,
which acquire a four-momentum (squared) dependence in the case of virtual
photons
$q^2\neq 0$.

By the late 1990s there is nearly uniform consensus in the physics community
that
indeed there exist a small quadrupole components in the electromagnetic
$N\Delta$
transition \cite{currexp,trento}. In the case of real photons one nowadays
believes an
Re$\left[EMR(0)\right]\approx-1/\%\ldots -4\%$, but a more precise
determination of this
fundamental
property of the nucleon has been surprisingly elusive and hotly contested
throughout
the past decade, both among theorists and among experimentalists. It is our
hope
that the ongoing experiments \cite{pap}
of electroproduction of $\Delta$(1232) with the resulting
better information on $q^2$-dependence of the $N\Delta$-transition form
factors,
as well as on EMR$(q^2)$, CMR$(q^2)$ will lead to clear picture of the
underlying
physics and enable us to identify the relevant degrees of freedom for several
different
regimes of momentum transfer. A dramatic change of the physics underlying the
electromagnetic $N\Delta$ transition is very much expected from {\em
perturbative} QCD,
which predicts for ``large'' $Q^2\equiv -q^2$ that
EMR$(Q^2)\rightarrow + 1$. At which finite $Q^2$ the crossover from a
negative to a
positive EMR should happen and whether this point is kinematically accessible
at
present/future electron scattering machines is a further issue of current
debate and interest, {\it e.g.} \cite{cm}.

>From a theorist's perspective,
treatments of the electromagnetic $N\Delta$ transition may be grouped
into two categories:
\begin{enumerate}
\item Calculations of the electromagnetic $\Delta N\gamma^\ast$ vertex with its
associated 3 complex form factors {\it per se},
using different theoretical frameworks that aspire to
fundamentally-based descriptions of the $N-\Delta$ system. The chiral Bag
model({\it e.g.}~\cite{XBM}), quark models with meson exchange currents
({\it e.g}~\cite{Buch}), lattice gauge theory ({\it e.g.}~\cite{Derek})
and effective chiral lagrangians ({\it e.g.}~\cite{Butler,Lucio}) constitute
such attempts.
\item
      Direct theoretical treatments~\cite{th,disp} of the full scattering
 processes ({\em e.g.}
$e N \rightarrow  e^\prime N \pi,\, e N\rightarrow e^\prime N \gamma$) in the
$\Delta$(1232) resonance region, either based on phenomenological Lagrangians
supplemented with a method of choice to unitarize the amplitudes or
dispersion relations. A point of strong contention therein is the issue of
separation of background vs. resonance contributions. For
a recent summary of the status of the resulting EMR/CMR extractions we refer to
the
talk by Workman~\cite{Work}.
\end{enumerate}

In the present work we calculate the $\Delta$(1232) to  $N$ radiative
transition $\Delta\rightarrow N \gamma^\ast$ in the region of small ({\it i.e.}
$Q^2<0.2$
GeV$^2$) photon virtuality, utilizing a recently developed
effective chiral lagrangian approach  \cite{short,long} that
systematically incorporates the spontaneous and the explicit breaking of the
chiral symmetry of QCD. A small scale $\varepsilon =
\{p,m_\pi,\delta\}$ denoting, collectively, small momenta, the
pion mass and the Delta-nucleon mass splitting
 is used to establish a systematic power-counting, thus telling us
precisely which diagrams/vertices have to be included if we want to calculate
up
to a certain order in $\epsilon$.
This approach allows for an efficient inclusion of $\Delta$(1232) degrees of
freedom
consistent with the underlying chiral symmetry of QCD and is referred to as the
``Small Scale Expansion'' (SSE) \cite{short,long}, constituting a
phenomenological
extension of Heavy Baryon Chiral Perturbation Theory \cite{BKM}.
The formalism can be used to {\it calculate}  both the vertex as well as
full scattering amplitudes--here we focus on the former. Clearly, as we are
dealing
with a {\it low energy} effective
theory, the $q^2$ dependence of a given physical quantity can be trusted
only at low values\footnote{Similar calculations
using SSE have been performed for the electromagnetic form factors of the
nucleon
and good agreement with experimental data has been found in the
$Q^2<0.2$GeV$^2$
regime \cite{talk,BFHM}.}. Accordingly, our final applications will be
discussed in this
spirit.

The most general form of the $\Delta\rightarrow N\gamma^\ast$ radiative
decay amplitude that complies with Lorentz covariance,
gauge invariance and parity conservation is described by three
form factors. We begin our discussion from the widely used form\footnote{
Equivalent forms, obtained via use of the equations
of motion, can be found in the vast literature on this subject.}
\begin{eqnarray}\label{eq1}
i{\cal M}_{\Delta\rightarrow N\gamma}^{rel}
=&-\frac{e}{2M_N} \bar{u}(p_N) \gamma_5 &\left[
   g_1(q^2)( {\thru q} \epsilon_\m - {\thru \e} q_\m )
 + {g_2(q^2)\over 2M_N} (p_N\dt\e\, q_\m - p_N\dt q\, \e_\m) \right.\nonumber
\\
& &\left.
 + {g_3(q^2)\over 2 M_N} (q\dt\e\, q_\m - q^2 \e_\m)\right]u^\m_\D(p_\Delta) \
.
\end{eqnarray}
Here $M_N$ is the nucleon mass, $p_{N,\Delta}^\mu$ denotes the relativistic
four-momentum of the nucleon, Delta
and $q_\mu$, $\epsilon_\mu$ are the photon momentum
and polarization vectors, respectively. The Delta is described via a
Rarita-Schwinger spinor $u_\Delta^\mu(p_\Delta)$ with free Lorentz index $\mu$.
The $M1$, $E2$ and $C2$ multipoles allowed in
${3\over 2}^+\rightarrow{1\over 2}^+$ electromagnetic transitions
can be cast as linear combinations of the form factors
$g_1, g_2$ and $g_3$~\cite{definition}.

In this work
we study the radiative vertex to ${\cal O}(\varepsilon^3)$ in the above
mentioned
SSE formalism. This constitutes the
first order where pion-nucleon and pion-delta loop graphs enter the vertex.
As will be shown later, an ${\cal O}(\varepsilon^3)$ calculation
entails corrections up to ${\cal O}(1/\Lambda_\chi^2)$, with
$\Lambda_\chi$ the chiral symmetry breaking scale. One of the main tasks
during this calculation has been the consistent matching between the
results of our
perturbative calculation and the most general vertex parameterization
as given for
example in Eq.(\ref{eq1}). In order to provide for a stringent test of
our
new predictions with experiment we note that a complementary calculation
of the full pion photoproduction amplitude in the $\Delta$(1232) region
to the same order in $\epsilon^3$ is in progress \cite{photo}.

There have been previous analyses of the radiative transition in  a similar
theoretical approach by Butler et al.~\cite{Butler} and  by Napsuciale and
Lucio~\cite{Lucio}. Our work differs from the aforementioned references in
the following aspects:
\begin{itemize}
\item The most crucial difference is that we address the {\it form factors}
      and not just the real-photon point. This entails (a) a $q^2$
      dependence in our expressions and (b) the presence of an additional
      form factor ($g_3$), or, equivalently, our calculation yields
      the CMR$(q^2)$, in addition to the EMR$(q^2)$.
\item
     SSE systematically keeps track of $1/M$ (i.e., relativistic)
corrections to lower order couplings. These corrections have not been
included in previous analyses \cite{Butler,Lucio}. We also find that our
identification of the form factors differs from the one used
in \cite{Lucio} at the real photon point.
\end{itemize}

This article is structured as follows. In the next Section we
briefly review the essentials of the ``small scale expansion'' and its
power counting.
In Section III, we discuss the most general form of the Pauli-reduced
transition amplitude which is consistent with the ${\cal
O}(\varepsilon^3)$ calculation in the SSE. We then proceed, in
Section IV, to calculate the loop as well as chiral counterterm
contributions. The resulting expressions for the form factors are
identified in Section V, while those for the multipoles in Section VI.
Numerical values for the $EMR(q^2),\;CMR(q^2)$ using presently available
(theoretical)
input are furnished in Section VII. In the same Section we give numerical
as
well as analytical results for (complex) slopes of the three
form factors. We summarize and offer our perspective for future efforts in
the concluding Section. Finally, we devote an Appendix to the discussion
of technical matters.

\section{Chiral
Lagrangians and the ``Small Scale Expansion''}

\subsection{Heavy Baryon ChPT}\label{s2}

QCD being a strongly coupled theory at low (of the order of 1 GeV)
energies, renders traditional perturbative expansions in the coupling
constant inadequate. Chiral perturbation theory offers an alternative
perturbative expansion, namely one that is realized in terms of the
external momenta involved in a given physical process. The original
strategy was based on the notion that at low energies
an effective theory of QCD will involve only the nearly massless
(i.e. the Goldstone bosons: pions etc.) degrees of freedom~\cite{Weinb,GLeut}.
Accordingly, chiral perturbation theory has been very successful with
respect to applications in the meson sector. With the inclusion of
baryons, however, the systematic power counting of ChPT fails, since
baryon masses $M_B (\ge 1 $ GeV) cannot join the set of the expansion
parameters $\{ {\mbox{external momenta}, m_\pi} \}$ as they are by no
means  small and remain finite in the chiral limit.

A systematic power counting can, nevertheless, be defined through a
splitting of the nucleonic field degrees of freedom into heavy-light modes
and integrating out the former. The cost of this procedure is to burden
the effective description with additional,  higher order contact
interactions. Generalizing recent developments in heavy quark effective
theories (see, e.g.,
\cite{WIsgur},\cite{Georgi}) heavy fermion methods were first applied to
baryon chiral
perturbation theory by Jenkins and Manohar~\cite{JenkManoh}.
The basic premise is the adoption of a non-relativistic
mode of description, which entails a restriction to four-velocities of the
form $v_\mu \sim (1+|{\pmb\delta}|,{\pmb\delta}),\,|{\pmb\delta }|
\ll 1$. On an operational level this means that all momentum dependence in the
theory
is only governed by (non-relativistic) soft momenta $k_\mu$, defined via
\begin{equation}
p_\mu\,=\,M_0\,v_\mu\,+\,k_\mu,
\end{equation}
where $p_\mu$ is a typical nucleon relativistic four-momentum and $M_0$
corresponds to
the nucleon mass in the chiral limit. The range of validity of the resulting
effective theory demands that each component $k_\mu<<\Lambda_\chi$, with the
chiral symmetry breaking scale being $\Lambda_\chi\approx1$GeV. This approach
is commonly referred to as Heavy Baryon Chiral Perturbation Theory (HBChPT).

In the construction of the effective theory we will follow the systematics laid
out in \cite{BKKM}. The general philosophy is to take the fully relativistic
theory as the starting point and then perform a systematic non-relativistic
reduction. This procedure automatically guarantees the proper $1/M$
corrections to the couplings in higher order
Lagrangian terms.
Alternatively, one could start
from a general Lagrangian within the non-relativistic framework, but then
has
to implement the so called ``reparameterization invariance''. The latter
approach is, for example, quite common in the field of heavy quark
effective
theories \cite{Finke}.

We now briefly sketch the derivation of the (non-relativistic) chiral
Lagrangians for matter fields. For details we refer the extensive
literature of
reviews ({\it e.g.} \cite{long},\cite{BKM},\cite{Ecker})

We start from the chiral relativistic SU(2) Lagrangian for nucleons
\begin{equation}
{\cal L}_N=\bar{\psi}_N\; \Gamma_N \; \psi_N \; , \label{eq:rellag}
\end{equation}
with the relativistic nucleon isospinor field
$\psi_N=\left(\begin{array}{c}\psi_p\\ \psi_n\end{array}\right)$ and
\begin{eqnarray}
\Gamma_N&=&\Gamma_N^{(1)}+\Gamma_N^{(2)}+\ldots \label{eq:gamma}
\end{eqnarray}
being a string of general nucleon-nucleon transition matrices $\Gamma_N^{(n)}$
of increasing chiral power $n$ \cite{GSS}. For example, to leading order one
obtains the well-known structure
\begin{eqnarray}
\Gamma_N^{(1)}&=&i{\thru{\cal D}}-M_0+\frac{\dot{g_A}}{2}\thru{u}\gamma_5 \; ,
\end{eqnarray}
where ${\cal D}_\mu$ denotes the chiral covariant
derivative. The parameter $\dot{g_A}$ corresponds to
the axial vector coupling constant (in the chiral limit) and the
chiral tensor $u_\mu$ describes the coupling of an odd number of pions with
nucleon. For more details we refer to \cite{GSS}.

The second step is a redefinition of the relativistic nucleon fields via
\begin{eqnarray}
N&=&e^{iM_0 v \cdot x}\;P_v^+\;\psi_N \nonumber \\
H&=&e^{iM_0 v \cdot x}\;P_v^-\;\psi_N \; ,
\end{eqnarray}
with the velocity-dependent projection operators
$P_v^\pm=\frac{1}{2}(1\pm\thru{v})$. $N$ is typically called the ``light''
field, whereas $H$ is commonly referred to as the ``heavy'' field. The
relativistic Lagrangian Eq.(\ref{eq:rellag}) then takes the general form
\begin{equation}\label{ck4}
{\cal L}_N\,=\,\bar{N}{\cal A}_NN+(\bar{H}{\cal B}_NN+h.c.)-\bar{H}{\cal
C}_NH,
\end{equation}
where the matrices ${\cal A}_N,{\cal B}_N,{\cal C}_N$ are defined via
\begin{eqnarray}
{\cal A}_N=P_v^+\Gamma_NP_v^+\; ,\quad {\cal B}_N=P_v^-\Gamma_NP_v^+\; ,\quad
{\cal C}_N=P_v^-\Gamma_NP_v^-\; .
\end{eqnarray}
We note that each matrix consists of an infinite string of terms of increasing
chiral power $n$, analogous to Eq.(\ref{eq:gamma})\footnote{Technically
speaking, matrices ${\cal A}_N,{\cal B}_N$ start with chiral power $n=1$, while
${\cal C}_N$ begins with $n=0:\; {\cal C}_N^{(0)}=2M_0$. This appearance of a
large mass term $2M_0 > \Lambda_\chi$ in ${\cal C}_N$ is also the reason why
the $H$ fields are denoted ``heavy'' and ultimately get integrated out.}.

Now one shifts the fields and integrates out the heavy component $H$. The
resulting (non-relativistic) effective Lagrangian reads
\begin{equation}\label{ck6}
{\cal L}^{eff}_N=\bar{N}({\cal A}_N+\gamma_0{\cal B}_N^\dagger\gamma_0{\cal
C}_N^{-1}{\cal B}_N)N   \ . \label{eq:leff}
\end{equation}
For the case of SU(2) the explicit form of this effective Lagrangian for
spin
1/2 nucleons has been worked out up to $n=3$ by several groups ({\em e.g.}
\cite{EM96,FMS,BFHM}). Generalizations to even higher orders are under way
\cite{priv}. The important point to note is that the inverse of matrix ${\cal
C}_N$ is calculated perturbatively, which confines the resulting effective
Lagrangian to the non-relativistic regime.

\subsection{The Small Scale Expansion}

HBChPT as described above, has been quite successful for
scattering processes off a single nucleon near
threshold (for reviews see \cite{BKM,ChPT97}). At higher energies,
however, the
contribution from nucleon resonances like $\Delta$(1232) can no longer be
parameterized via higher order nucleon-nucleon couplings in $\Gamma_N^{(2)},
\Gamma_N^{(3)}$, etc. At some point, i.e. once explicit propagation of a
nucleon- or meson-resonance has to be included, the above described contact
interaction approach breaks down. So if one is interested in kinematic
conditions of such dynamic resonance contributions, or in investigations
into the low energy
structure of nucleon resonances, it becomes mandatory to include low lying
resonances as explicit degrees of freedom in the effective Lagrangian
Eq.(\ref{eq:leff}). In particular this means the inclusion of the spin
3/2
nucleon resonance $\Delta$(1232) in the case of an SU(2) analysis.

The first efforts towards this direction were performed by Jenkins and
Manohar
\cite{JM2}. In the present work, however, we follow a specific generalization
of the
construction method of Ref.\cite{BKKM} (as outlined above), which is
called the
``small scale expansion'' ({\em SSE}) \cite{short,long,BFHM,kambor,Compton2}.
The main
difference to HBChPT lies in the fact that {\em the chiral power counting is
modified in a phenomenologically inspired fashion}. In the {\em SSE} approach
one expands in the small scale $\varepsilon=\{ \mbox{soft
momenta},m_\pi,\delta_0\} $, where
$\delta_0=\dot{M}_\Delta-M_0$ corresponds
to the Delta-nucleon
mass splitting in the chiral limit, whereas in HBChPT one expands in the
quantity
$p=\{ \mbox{soft momenta},m_\pi \} $\footnote{In strict HBChPT
$\delta_0$
counts as a quantity of order $p^0$. This is formally correct but can lead to
poor convergence properties in the perturbation series.
For more details we refer to the
discussion in \cite{Compton2} regarding the spin-polarizabilities of the
nucleon.}. The chiral power counting of HBChPT as an expansion of all
quantities in a power series governed by $p^n$ is taken over by {\em SSE} as an
expansion of all quantities in a power series governed by $\varepsilon^n$.
In the
following we only briefly discuss the construction of the relevant SU(2)
Lagrangians with explicit pion, nucleon and Delta degrees of freedom and
their
couplings to arbitrary external fields. For details we refer the interested
reader to \cite{short,long,BFHM}.

The starting point this time is a set of coupled relativistic SU(2)
Lagrangians
of relativistic nucleon and Delta fields $\psi_N,\psi_\mu$\footnote{For
notational simplicity we suppress all explicit isospin indices.}
\begin{eqnarray}
{\cal L}_N&=&\bar{\psi}_N\; \Gamma_N \; \psi_N \;  \nonumber \\
{\cal L}_\Delta&=&\bar{\psi}_\mu\; \Gamma_\Delta^{\mu\nu} \; \psi_\nu \;
 \nonumber \\
{\cal L}_{\Delta N}&=&\bar{\psi}_\mu\; \Gamma_{\Delta N}^\mu \; \psi_N \; + \;
h.c. \; ,
\label{eq:coupled}
\end{eqnarray}
where each baryon matrix $\Gamma_X,\;X=N,\Delta,\Delta N$ is a string of terms
of increasing power in $\varepsilon^n$, analogous to Eq.(\ref{eq:gamma}).
For the spin 3/2, isospin 3/2 Delta field we utilize a
Rarita-Schwinger-isospurion notation, as laid out in the appendices of
Ref.\cite{long}. In the Rarita-Schwinger formulation for spin 3/2 fields one
encounters the well-known redundancy of degrees of freedom, which calls for six
projection operators before one is able to isolate the ``light'' spin 3/2
component $T_\mu$ from the ``heavy'' one $G_\mu$:
\begin{eqnarray}\label{ck8}
T_\mu&=&e^{iM_0 v\cdot x}\; P_v^+\;P_{(33)\mu\nu}^{3/2}\;\psi^\nu\; ,
\end{eqnarray}
with
\begin{equation}\label{ck9}
P_{(33)\mu\nu}^{3/2}=g_{\mu\nu}-{1\over 3}\gamma_\mu\gamma_\nu- {1\over 3}
({\thru v}\gamma_\mu v_\nu+v_\mu\gamma_\nu{\thru v}).
\end{equation}
The ``heavy'' spin 3/2 field $G_\mu$ is a five-component object. An explicit
representation can be found in \cite{long} but is not needed here.

Suppressing all spin-isospin indices, the coupled set of relativistic
Lagrangians in Eq.(\ref{eq:coupled}) can then be written as
\begin{eqnarray}
{\cal L}_N&=&\bar{N}{\cal A}_NN+(\bar{H}{\cal B}_N N+h.c.)-\bar{H}{\cal
C}_N H \nonumber \\
{\cal L}_\Delta&=&\bar{T}{\cal A}_\Delta T+(\bar{G}{\cal B}_\Delta T
+h.c.)-\bar{G}{\cal C}_\Delta G \nonumber  \\
{\cal L}_{\Delta N}&=&\bar{T}{\cal A}_{\Delta N}N+\bar{G}{\cal
B}_{\Delta N}N+\bar{H}{\cal D}_{N\Delta}T+\bar{H}{\cal
C}_{N\Delta}G+h.c.\label{ck11}
\end{eqnarray}
Now one integrates out the ``heavy'' nucleon and ``heavy'' Delta
components
$H,G$ and finds the resulting (non-relativistic) effective Lagrangian with
explicit nucleon and Delta fields:
\begin{equation}\label{ck12}
{\cal L}_{\em SSE}^{eff}=\bar{T}\tilde{{\cal A}}_\Delta
T+\bar{N}\tilde{{\cal A}}_NN+[\bar{T}
\tilde{{\cal A}}_{\Delta N}N+h.c.],
\end{equation}
with
\begin{eqnarray}\label{ck13}
\tilde{{\cal A}}_\Delta & =  & {\cal A}_\Delta+\gamma_0
\tilde{{\cal D}}^\dagger
_{N\Delta}\gamma_0\tilde{{\cal C}}^{-1}_N\tilde{{\cal D}}_{N\Delta}
+\gamma_0{\cal B}^\dagger_\Delta\gamma_0{\cal C}^{-1}_\Delta
 {\cal B}_\Delta \nonumber\\
\tilde{{\cal A}}_N &=& {\cal A}_N+\gamma_0\tilde{{\cal B}}^\dagger_N
\gamma_0\tilde{{\cal C}}^{-1}_N\tilde{{\cal B}}_N+\gamma_0
{\cal B}^\dagger_{\Delta N}\gamma_0{\cal C}_\Delta^{-1}{\cal B}_{\Delta N}
\nonumber\\
\tilde{{\cal A}}_{\Delta N}&=&{\cal A}_{\Delta N}+\gamma_0
\tilde{{\cal D}}^\dagger_{N\Delta}\gamma_0\tilde{{\cal C}}^{-1}_N\tilde
{{\cal B}}_N+\gamma_0{\cal B}^\dagger_\Delta\gamma_0{\cal C}_\Delta
^{-1}{\cal B}_{\Delta N}
\end{eqnarray}
and
\begin{eqnarray}\label{ck14}
\tilde{{\cal C}}_N&=&{\cal C}_N-{\cal C}_{N\Delta}{\cal C}_\Delta^{-1}
\gamma_0{\cal C}^\dagger_{N\Delta}\gamma_0 \nonumber\\
\tilde{{\cal B}}_N&=&{\cal B}_N+{\cal C}_{N\Delta}{\cal C}_\Delta^{-1}
{\cal B}_{\Delta N} \nonumber\\
\tilde{{\cal D}}_{N\Delta}&=&{\cal D}_{N\Delta}+{\cal C}_{N\Delta}{\cal
C}_\Delta^{-1}{\cal B}_\Delta  \ .
\end{eqnarray}

The explicit form of the effective Lagrangian Eq.(\ref{ck12}) in SSE has
been worked out to $n=2$ in Ref.\cite{long}. For $n=1$
one finds
\begin{eqnarray}\label{Lagrangian}
\tilde{{\cal A}}_{N}^{(1)}={\cal A}_{N}^{(1)}&=&  i v\cdot D + \dot{g}_A S\cdot
u
                           \nonumber\\
\tilde{{\cal A}}_{N\D}^{(1)}={\cal A}_{N\D}^{(1)}&=&\dot{g}_{\pi N\D} \; w^i_\m
                            \nonumber\\
\tilde{{\cal A}}_{\D}^{(1)}={\cal A}_{\D}^{(1)}&=& -\left( i v\cdot D^{ij}
-
\delta_0
                            \xi^{ij}_{3/2} +\dot{g}_1\; S\cdot u^{ij} \right)
                            g_{\mu\nu}  \ .
\end{eqnarray}
In addition to the structures already discussed in section \ref{s2} one
encounters the
Pauli-Lubanski spin vector $S_\mu$ and  the $\pi N\Delta,\pi\Delta\Delta$ bare
coupling parameters $\dot{g}_{\pi N\Delta},\dot{g}_1$. Furthermore, the
chiral
tensors $w_{\mu}^i,u_\mu^{ij}$ parameterize the coupling of an odd number of
pions to an $N\Delta$, and a $\Delta\Delta$ transition current, whereas
$D_\mu^{ij}$
corresponds to the chiral covariant derivative acting on a spin 3/2 - isospin
3/2 field. With $i, j$ = 1, 2, 3 we denote explicit isospin indices and
$\xi^{ij}_{3/2} = {2\over 3}\delta^{ij} - {i\over 3} \epsilon^{ijk} \tau^k$
corresponds
to the
isospin 3/2 projector. For more details we refer to \cite{long}.

At $n=2$ we are only interested in $\Delta N\gamma$ vertices, as will be
discussed in section \ref{born}.
The relevant relativistic ${\cal O}(\epsilon^2)$ $\Delta N\gamma$
contact terms can
be parameterized in terms of the low energy constants (LECs) $b_1$, $b_6$ as
follows
\begin{equation}\label{rel1}
{\cal L}^{(2)}_{\Delta N\gamma} = {i b_1\over 4 M_0} \bar\psi^\mu_i
\left[ g_{\mu\nu}
+ y_1 \gamma_\mu\gamma_\nu \right] \gamma_\rho\g5
f^{\nu\rho}_{i\hspace{0.01in}+} \psi _N-\frac{b_6}{2M_0^2} \bar\psi^\mu_i
\left[ g_{\mu\nu}
+ y_6 \gamma_\mu\gamma_\nu \right] \g5
f^{\nu\rho}_{i\hspace{0.01in}+} D_\rho\psi _N+h.c.
 \ ,
\end{equation}
where $f^{\rho\n}_{i\hspace{0.01in}+} \equiv  {1\over 2}
 {\rm Tr} \left[ f^{\rho\nu}_+ \tau^i\right]$ denotes
the isovector photon component and $y_{1,6}$ are the
so-called off-shell coupling constants.

Performing the transition to the heavy baryon fields for the vertices of
interest to
our calculation one finds~\cite{long}
\begin{eqnarray}
\tilde{{\cal A}}_{N\D}^{(2)}&=&-i\;\frac{b_1}{2M_0}\;f_{+\mu\nu}^iS^\nu +
\dots \, ,
\label{eq:a2}
\end{eqnarray}
which is independent of the couplings $y_1,b_6,y_6$.
In the radiative decay of the $\Delta$(1232) the
fact that there is no S-wave multipole allowed precludes the possibility
of having an ${\cal O}(\epsilon)$ $\Delta N\gamma$ vertex.
The parameter $b_1$ therefore denotes the leading $\gamma N\Delta$ coupling and
carries the main strength of the M1 $\Delta N\gamma$
transition \cite{short,long,Compton2}.

While the desired analysis for $n=2$ is quite simple, the situation at
$n=3$ is more complex. In fact,
there is no agreement in the literature about number and/or structure
of the subleading $\Delta N\gamma$ vertices ({\em e.g.} compare refs.
\cite{Butler,Lucio,th}). In the next subsection we shall
explicitly construct the $n=3$ $\Delta N\gamma$ vertices according to
the SSE approach.

\subsection{Construction of ${\cal O}(\varepsilon^3)$ Vertices}

Referring to the SSE master-formula
Eqs.(\ref{ck12}-\ref{ck14}), the relevant Lagrangian
${\cal L}^{(3)}_{\Delta
 N\gamma}$  contains terms which have two separate origins---the
dimension six contact interactions contributing to
$A_{N\Delta}^{(3)}$ and possible $1/M$ corrections related to the ${\cal
O}(\varepsilon^2)$ $\Delta
N\gamma$ vertices. The latter are given by
\begin{equation}\label{corr}
        L_{\Delta\rightarrow N\gamma^\ast}^{(3)\, fixed}=
        \bar N(\gamma^0 {\cal B}_N^{\dagger(1)}\gamma^0
             {\cal C}_N^{(0)\hspace{0.01in}-1}
          {\cal D}_{N\Delta}^{(2)}
          + \gamma_0 {\cal B}_{\Delta N}^{\dagger(2)}\gamma_0
          {\cal C}_{\Delta}^{(0)\hspace{0.01in}-1}
           {\cal B}_{\Delta}^{(1)})T \ ,
\end{equation}
where the superscripts denote the SSE dimension. The pertinent coupling
matrices can be found in \cite{long}, {\it except} for the structures
denoted as ${\cal B}_{\Delta N}^{\dagger(2)}$, ${\cal D}_{N\Delta}^{(2)}$
which only start contributing at ${\cal O}(\varepsilon^3)$. We therefore
construct these matrices, but only for the vertices of interest in our
calculation. Using Eq.(\ref{rel1}) and the orthogonality properties of the
light-heavy
projection operators \cite{long} we find (dots signify terms that are
irrelevant to
the present analysis)
\begin{equation}\label{relfirst}
{\cal D}_{N\D}^{(2)} = \frac{i(b_1+2b_6)}{4M_0} P_-
P_{(33)}^{3/2\hspace{0.01in}\n\m}
\gamma_5 f_{\nu\rho}^+ v^\rho+...\equiv {\cal D}_{N\D \gamma}^{(2)}+...
\end{equation}
and (for the $b_1$ part)
\begin{equation}\label{rel2}
{\cal B}_{\Delta N}^{(2)b_1} = {i b_1\over 4 M_0}
\left[ \begin{array}{c}
-P_{(33)}^{3/2\hspace{0.01in}\mu\nu} v^\rho \\
 (1+3y)P_{(11)}^{1/2\hspace{0.01in}\mu\nu} (\gamma^\rho+v^\rho) -\sqrt{3}
y
P_{(12)}^{1/2\hspace{0.01in}\mu\nu} v^\rho \\
-(1+ y)P_{(22)}^{1/2\hspace{0.01in}\mu\nu} v^\rho   +\sqrt{3} y
P_{(21)}^{1/2\hspace{0.01in}\mu\nu} (\gamma^\rho+v^\rho) \\
-(1+3y)P_{(11)}^{1/2\hspace{0.01in}\mu\nu} v^\rho   +\sqrt{3} y
P_{(12)}^{1/2\hspace{0.01in}\mu\nu} (\gamma^\rho+v^\rho) \\
 (1+ y)P_{(22)}^{1/2\hspace{0.01in}\mu\nu} (\gamma^\rho+v^\rho) -\sqrt{3}
y
P_{(21)}^{1/2\hspace{0.01in}\mu\nu} v^\rho \\
\end{array}\right]
\g5 f_{\nu\rho}^+ \ ,
\end{equation}
where we have suppressed isospin indices.
In a similar manner one obtains
the contribution to $B_{\Delta N}^{(2)}$ associated with
$b_6$. The complete 1/M corrections at ${\cal O}(\epsilon^3)$ which had to
be constructed via Eq.(\ref{corr}) then read
\begin{eqnarray}\label{rel3}
L_{\Delta\rightarrow N\gamma^\ast}^{(3)\, fixed}&=&
{b_1 + 2b_6 \over 4 M^2_0} \bar N (S\dt \stackrel{\scriptstyle
\leftarrow}{D}) v^\rho f^i_{\mu\rho +}T^\mu_i
+\frac{b_1-2b_6}{4M_0^2} \bar N f_{\rho\mu +}^k
v^\rho\xi^{kj}_{3/2} S\cdot D^{ji}T^\mu_i  \nonumber \\
& &+\frac{b_1}{2M_0^2} \bar Nf^k_{\rho\beta +}S^\rho v^\beta\xi_{3/2}^{kj}
D_\mu^{ji}T^\mu_i.
\end{eqnarray}
Interestingly, one observes that the off-shell couplings $y_1,y_6$ drop out
in the final result.

Finally we have to determine the dimension six $\gamma N\Delta$ counterterms
contained
in $A_{N\Delta}^{(3)}$ as mandated by Eq.(\ref{ck12}). Once more we have to
start
from that part of the most general relativistic ${\cal O}(\epsilon^3)$
lagrangian that
contains all allowed $\gamma N\Delta$ vertices at this order, in which the
chiral tensor
$f_{\nu\alpha\beta}^{i+}\equiv\frac{1}{2}{\rm Tr}\left\{ \tau^i
\left[D_\nu,f_{\alpha\beta}^+\right] \right\} $ plays a dominant part.
At a first glance the following structures are possible candidates:
     $
     g^{\mu\nu}\sigma^{\alpha\beta}\gamma_5f_{\nu\alpha\beta}^{i+}$,
     $
     g^{\mu\nu}g^{\alpha\beta}\gamma_5f_{\nu\alpha\beta}^{i+}$,
     $
     g^{\mu\alpha}\sigma^{\nu\beta}\gamma_5f_{\nu\alpha\beta}^{i+}$,
     $
     g^{\mu\alpha}g^{\nu\beta}\gamma_5f_{\nu\alpha\beta}^{i+}$,
     $\epsilon^{\nu\alpha\beta\mu}f_{\nu\alpha\beta}^{i+}$, where the
free index $\mu$ is to be contracted with the corresponding index of the
Rarita-Schwinger
Delta spinor in the initial state.
It is straightforward to verify that only the first and third quantities
will survive in $A_{N\Delta}^{(3)}$ and can be shown to give
identical contributions up to higher
order terms.  Accordingly we find
\begin{equation}\label{3op}
A_{N\Delta}^{(3)}=\frac{D_1}{4 M_{0}^2}
\;g^{\mu\nu}v^\alpha S^\beta
\frac{1}{2}{\rm Tr}\left\{ \tau^i
\left[D_\nu,f_{\beta\alpha}^+\right] \right\}
+\frac{i E_1}{2 M_0}\; \frac{\delta_0}{M_0}
 \;f_{+\mu\nu}^{\; i}\;S^\nu+...\equiv A_{N\Delta\gamma}^{(3)}+...,
\end{equation}
where $D_1$,  $E_1$ are now identified as the 2 $\gamma N\Delta$
low energy constants of ${\cal O}(\epsilon^3)$.
\footnote{The second term of
Eq.(\ref{3op}), proportional to $E_1$, appears in the ``small scale
expansion''
due to
the fact that
the Delta-nucleon mass splitting $\delta_0$ counts as a quantity of
${\cal O}(\epsilon)$. Therefore one has to construct additional
relativistic contact terms which are products of lower order interactions
and the new small scale $\delta_0$. For a discussion of this issue see
Ref.~\cite{BFHM}.}
This set is sufficient for
the renormalization of the process $\Delta->N\gamma^*$, a more detailed
discussion of
the construction of these counterterms can be found in \cite{photo}

Gathering the above results we determine the relevant SSE $\gamma N\Delta$
lagrangian
to ${\cal O}(\epsilon^3)$:
\begin{eqnarray}
{\cal L}_{\Delta N\gamma}^{(3)}&=&\bar{N}\; {\cal A}_{N\Delta\gamma}^{(3)}
\;T+\bar{N}\left[\gamma^0
{\cal B}_N^{\dagger(1)}\gamma^0\left( {\cal C}_N^{(0)}\right)^{-1}
{\cal D}_{N\Delta\gamma}^{(2)}
+ \gamma_0 {\cal B}_{\Delta N\gamma}^{\dagger(2)}\gamma_0\left( {\cal
C}_{\Delta}^{(0)}
\right)^{-1} {\cal B}_{\Delta}^{(1)} \right] T  +h.c.\nonumber \\
&=&\frac{D_1}{4 M_{0}^2}\; \bar{N}g^{\mu\nu}v^\alpha S^\beta
f_{\nu\beta\alpha}^{i\,+}
T_{\mu}^i+
\frac{i E_1}{2 M_0}\; \frac{\delta_0}{M_0}
\bar{N} \;f_{+\mu\nu}^{\; i}\;S^\nu \;T_{i}^\mu \nonumber \\
& &-{b_1 + 2 b_6 \over 4 M^2_0} \bar N (S\dt D) v^\rho
f_{\mu\rho}^{i+}T_i^\mu
+\frac{b_1-2b_6}{4M_0^2} \bar N f_{\rho\mu +}^k
v^\rho\xi^{kj}_{3/2} S\cdot D^{ji}T^\mu_i  \nonumber \\
& &+\frac{b_1}{2M_0^2} \bar Nf^k_{\rho\beta +}S^\rho v^\beta\xi_{3/2}^{kj}
D_\mu^{ji}T^\mu_i  +h.c.
\label{eq:lag3}
\end{eqnarray}

\section{ Non-relativistic reduction of the $\Delta\rightarrow N\gamma$ vertex}

Since the leading order Lagrangian counts as ${\cal O}(\varepsilon)$ and,
from the general exposition in the previous Section follows that
each new  order in the SSE contributes a power
$\varepsilon /M$,
the calculation of the radiative Delta
decay to ${\cal O}(\ve^3)$ will include all terms
up to ${\cal O}(1/M^2)$. Thus, in order to compare our calculation with
Eq.(\ref{eq1}) we must find the most general form of the
amplitude consistent with an ${\cal O}(1/M^2)$
calculation. In our case it is convenient to use the
$\Delta$(1232) rest frame where a Pauli reduction of the amplitude allows
one to identify
the various multipoles.

As far as Dirac spinors are concerned the situation with the Pauli
reduction is well-known. For the Delta field, on the other
hand, we think that is worth providing our conventions. We start from a
typical ({\it e.g.} \cite{EW})
Rarita-Schwinger representation of the $\Delta$(1232)
spinor with the subsidiary conditions
\begin{equation}
\gamma_{\mu}u_{\Delta}^{\mu}=0,
\hspace{0.7in}
\partial_{\mu}u_{\Delta}^{\mu}=0.
\end{equation}
A specific representation consistent with these constraints is the
following
\begin{equation}
u_{\Delta}^{\mu}=\left(\frac{E_{\Delta}+M_{\Delta}}{2M_{\Delta}}\right)
^{1/2}\left(\begin{array}{c}1 \cr\frac{ {\pmb\sigma}\cdot{\bf p}_
{\Delta}}{E_{\Delta}+M_{\Delta}} \end{array}\right)\Sigma^{\mu}
\chi_{3\over 2},
\end{equation}
with $E_{\Delta}$, $M_{\Delta}$ and ${\bf p}_{\Delta}$ the energy, mass
and three momentum of the Delta particle, respectively and
$\Sigma^{\mu}$ the four-vector given by
\begin{equation}
\Sigma^\mu = \left[
             \frac{{\bf S}\cdot{\bf p}_{\Delta}}{M_{\Delta}},
{\bf S}+\frac{{\bf S}\cdot{\bf p}_{\Delta}}
{M_{\Delta}(E_{\Delta}+M_{\Delta})}\ {\bf p}_{\Delta}\right],
\end{equation}
where $\bf S$ denotes the spin $3/2 \rightarrow 1/2$ transition
matrices~\cite{EW}.

Next, we have to expand the various terms entering Eq.(\ref{eq1}) in
powers of $1/M$ and
restrict ourselves up to ${\cal O}(1/M^2)$ accuracy. Using the analogue of
the
four-component notation of the SSE
\begin{eqnarray}
u_v(r_N)&=&P_v^+\; u(p_N) \nonumber \\
u_{v,\Delta}^\mu(r_\Delta)&=&P_v^+\; u_\Delta^\mu(p_\Delta),
\end{eqnarray}
with $p^\mu_X=M_0 v^\mu+r^\mu_X,\; X=N,\Delta$,
we finally arrive at the following expression
\begin{eqnarray}\label{exp}
i{\cal M}_{\Delta\rightarrow N\gamma}^{(1)} =& \sqrt{2\over 3}e \
\bar{u}_v(r_N)&\left\{
 (S\cdot \epsilon) q_\mu  \
\left[ \frac{g_1(q^2)}{M_N}+{\cal O}(1/M_N^3)  \right] \right.
\nonumber \\
& &+ (S\cdot q) \epsilon_\mu  \
\left[
-\frac{g_1(q^2)}{M_N}-\frac{\delta}{2M_N^2}g_1{(0)}+\frac{\delta}{4M_N^2}
g_2{(0)}+{\cal O}(1/M_N^3) \right]
\nonumber \\
& &+ (S\cdot q) (v\cdot\epsilon) q_\mu  \
\left[ \frac{( g_1{(0)} - {1\over 2} g_2{(0)} )}{2M_N^2} +{\cal
O}(1/M_N^3) \right]
\nonumber \\
& &\left. + (S\cdot q) (q \cdot \epsilon) q_\mu  \
\left[ 0+ {\cal O}(1/M_N^3) \right]
\right\} u_{v,\Delta}^{\mu}(0)\ ,
\end{eqnarray}
which has been evaluated in the rest-frame of the Delta particle.

Some comments regarding the above equation are appropriate:
\begin{enumerate}
\item
      To the order we are working the energy $\omega$ of the outgoing
      photon four-momentum is $\omega = M_\Delta - \sqrt{M_N^2+{\bf q}^2}$ =
      $\delta -\frac{{\bf q}^2}{2M_N}+$ . . . =$\delta$ + $O(1/M_N)$.
\item
      For $q^2=0$ Eq.(\ref{exp}) is not in agreement with the analogous
      expression
      of Ref. \cite{Lucio}. There, the nucleon mass dependence of the
Dirac
      spinors
      apparently was not taken into account.
\item
      Eq.(\ref{exp}) has been derived under the assumption that
      all the form factors are quantities of ${\cal O}(\varepsilon^0)$ at
leading order.
\item
      It is
      inconsistent with
      our subsequent loop results. This can
      be most easily seen by the
      structure proportional to
      $\epsilon\cdot q$. According to the $1/M$ expansion this structure
      should not exist to ${\cal O}(\varepsilon^3)$.
      However, as it will turn out,
      the loop diagrams of Fig. 1 produce a non-trivial $q^2$ dependence
      which scales as $1/\Lambda_\chi^2 \sim 1/M^2$.
\end{enumerate}

We conclude that the popular form of the
relativistic  isovector nucleon-Delta transition current as given in
Eq.(\ref{eq1}), is not ideally suited for microscopic calculations of
these form factors which rely on $1/M$ expansions involving a chiral
power counting. We find it, therefore,
advantageous\footnote{The only alternative would be a $g_3$ form factor which
behaves as ${\cal O}(\epsilon^{-1})$ at $q^2=0$, {\em i.e.}
$g_3(0)\sim\frac{\Lambda_\chi}{m_\pi},\,\frac{\Lambda_\chi}{\delta}$. However,
such
a result would violate the
basic premise of non-relativistic formalisms like HBChPT or SSE that the large
scale $\Lambda_\chi \approx M_N, 4\pi F_\pi$ should {\it always} appear as
$\Lambda_\chi^{-n},\, n\geq 0$ in
order to ensure a consistent power counting.}, in view of the above remarks, to
define a new set of form factors, by rescaling the previous ones.
We propose the following definition
\begin{eqnarray}\label{fformf}
G_1(q^2)\equiv g_1(q^2), \quad G_{2}(q^2)\equiv g_{2}(q^2),
\quad G_{3}(q^2)\equiv{\delta \over M_N}g_{3}(q^2),
\end{eqnarray}
which leads to the relativistic $\Delta\rightarrow N\gamma^\ast$ transition
matrix element
\begin{eqnarray}\label{eq1neu}
i{\cal M}_{\Delta\rightarrow N\gamma}^{rel}
=& \sqrt{\frac{2}{3}}e\;\bar{u}(p_N) \gamma_5& \left[
   \frac{G_1(q^2)}{2 M_N}\left( {\thru q} \epsilon_\m - {\thru \e} q_\m \right)
 + \frac{G_2(q^2)}{4 M_N^2} \left(p_N\dt\e\, q_\m - p_N\dt q\,
\e_\m\right)
 \right. \nonumber \\
& &\left.
 + \frac{G_3(q^2)}{4 M_N\delta } (q\dt\e\, q_\m - q^2 \e_\m)\right]
u^\m_\D(p_\Delta) \ .
\end{eqnarray}
Eq.(\ref{eq1neu}) will serve as the reference point for future comparisons
with other calculations.
The advantage offered by  this parameterization can be seen best when
performing
a $1/M$ expansion of Eq.(\ref{eq1neu}), once more restricted to
${\cal O}(1/M^2)$:
\begin{eqnarray}\label{exp2}
i{\cal M}_{\Delta->N\gamma}^{NR} =& \sqrt{\frac{2}{3}} e \
\bar{u}_v(r_N)&\left\{
 (S\cdot \epsilon) q_\mu
\left[ \frac{G_1(q^2)}{M_N}+{\cal O}(1/M_N^3)  \right] \right.
\nonumber \\
& &+ (S\cdot q) \epsilon_\mu
\left[ -\frac{G_1(q^2)}{M_N}-\frac{\delta G_1(0)}{2M_N^2}+\frac{\delta
G_2(q^2)}{4M_N^2}
+\frac{q^2}{4M_N^2\delta}G_3(q^2)+{\cal O}(1/M_N^3) \right] \nonumber \\
& &+ (S\cdot q) (v\cdot\epsilon) q_\mu
\left[ \frac{G_1(0)}{2M_N^2} - \frac{G_2(q^2)}{4M_N^2}
+{\cal O}(1/M_N^3) \right] \nonumber \\
& &\left. + (S\cdot q) (q \cdot \epsilon) q_\mu
\left[ -\frac{G_3(q^2)}{4M_N^2\delta}
+ {\cal O}(1/M_N^3) \right]
\right\} u_{v,\Delta}^{\mu}(0)\ .
\end{eqnarray}
The above formula serves as the basic connection between our
(non-relativistic) results calculated via
{\em SSE} and the desired (relativistic) transition current Eq.(\ref{eq1neu}).
One can easily see that to ${\cal O}(\epsilon^3)$
one is sensitive to the leading and subleading behavior of the form factor
$G_1(q^2)$,
whereas the form factors $G_2(q^2), G_3(q^2)$ start out at $1/M^2$ and
therefore
only allow us to determine them to
leading order. One can also
check that Eq.(\ref{exp2}) is gauge-invariant to
${\cal O}(1/M^2)$ and allows for a non-vanishing $q^2$-dependent
contribution
in each of the four independent structures, consistent with the
{\em SSE} calculation to ${\cal O}(\varepsilon^3)$. We now move to the
details of the calculation.

\section{The calculation}
\subsection{Born Contributions}\label{born}

In the present subsection we discuss the Born contribution to the
amplitude. The relevant expressions
resulting from each $\Delta N\gamma^\ast$ tree level vertex have the following
explicit forms\footnote{To the order we are working we can identify the
parameters $M_0,\delta_0$ with the physical values
$M_N,\delta=M_\Delta-M_N$.}
\begin{eqnarray}\label{Mid}
i{\cal M}^{b_1}_{\Delta\rightarrow N\gamma^*}=& -\sqrt{\frac{2}{3}}
e\;\bar{u}_v(r_N)
&\left[
S\cdot q\;\epsilon_\mu\frac{b_1}{2M_N}-S\cdot \epsilon\;
q_\mu\frac{b_1}{2M_N}\right]u_{v,\Delta}^{\mu}(0) \nonumber \\
i{\cal M}^{E_1}_{\Delta\rightarrow
N\gamma^*}=& -\sqrt{\frac{2}{3}}  e\;\bar{u}_v(r_N) &\left[(S\cdot
\epsilon\;q_\mu-S\cdot q\;
\epsilon_\mu) \frac{E_1\delta}{2M_N^2}\right]u_{v,\Delta}^{\mu}(0)
\nonumber \\
i{\cal M}^{D_1}_{\Delta\rightarrow N\gamma^*}=& -\sqrt{\frac{2}{3}}
e\;\bar{u}_v(r_N)
&\left[
-S\cdot\epsilon\;q_\mu\frac{D_1\delta}{4M_N^2}+\epsilon_0\;S\cdot
q\;q_\mu\frac{D_1}{4M_N^2}\right]u_{v,\Delta}^{\mu}(0) \nonumber \\
i{\cal M}^{b_1\hspace{0.05in}rc}_{\Delta\rightarrow N\gamma^*}=
& -\sqrt{\frac{2}{3}}  e\;\bar{u}_v(r_N) &\left[S\cdot q\;
\epsilon_\mu\frac{b_1\delta}{4M_N^2}
+\epsilon_0\;S\cdot
q\;q_\mu\frac{-b_1}{4M_N^2}\right]u_{v,\Delta}^{\mu}(0) \nonumber \\
i{\cal M}^{b_6\hspace{0.05in}rc}_{\Delta\rightarrow N\gamma^*}=
& -\sqrt{\frac{2}{3}}  e\;\bar{u}_v(r_N) &\left[S\cdot q\;
\epsilon_\mu\frac{b_6\delta}{2M_N^2}
+\epsilon_0\;S\cdot
q\;q_\mu\frac{-b_6}{2M_N^2}\right]u_{v,\Delta}^{\mu}(0),
\end{eqnarray}
where the superscripts tabulate the relevant vertices, in relation to the
LECs, while $``rc"$ stands for ``relativistic correction".

We mention that,
the second relativistic correction Lagrangian term in Eq.(\ref{rel3})
gives {\it zero} contribution to our form factors. To see this, note
that the derivatives acting on the Delta
field  bring down factors $ (S\cdot r_\D)$ and $(r_\D\cdot T)$ where
$r_\D^\m$
is the $\Delta$(1232) soft momentum. In the $\Delta$(1232) rest frame,
where our calculation is performed one finds $r_\D^\m = \delta v^\mu$.
Both $(S\cdot r_\D)$ and $(r_\D\cdot T)$ therefore vanish
due to the light Delta constraints.
We do anticipate, on the other hand, that this operator will contribute
to the resonant Born diagram in pion photo-production where the
intermediate $\Delta$(1232) will also contain off-shell components of the Delta
field.

The overall Born contribution, therefore, is
\begin{eqnarray}
i{\cal M}_{\Delta\rightarrow
N\gamma^\ast}^{Born}&=&-\sqrt{\frac{2}{3}}e\;\bar{u}_v(r_N)\left[
S\cdot\epsilon\;q_\mu\left(\frac{-b_1}{2M_N}+\frac{(2E_1-D_1)\delta}{4
M_{N}^2}
\right)\right.\nonumber \\
& &+S\cdot q\; \epsilon_\mu\left(\frac{b_1}{2M_N}-\frac{E_1\delta}{2
M_{N}^2}+
\frac{(b_1+2b_6) \;\delta}{4 M_{N}^2}\right)
\nonumber \\
& &\left.+v\cdot\epsilon\;S\cdot q\;q_\mu\left(\frac{D_1}{4M_{N}^2}
-\frac{b_1+2b_6}
{4 M_{N}^2}\right)\right]u_{v,\Delta}^{\mu}(0).
\label{eq:born}
\end{eqnarray}

In the next subsection we will turn our attention to contributions from
loop corrections to the transition vertex.

\subsection{Loop contributions}

\begin{figure}[htb]
\begin{center}
\mbox{\epsfig{file=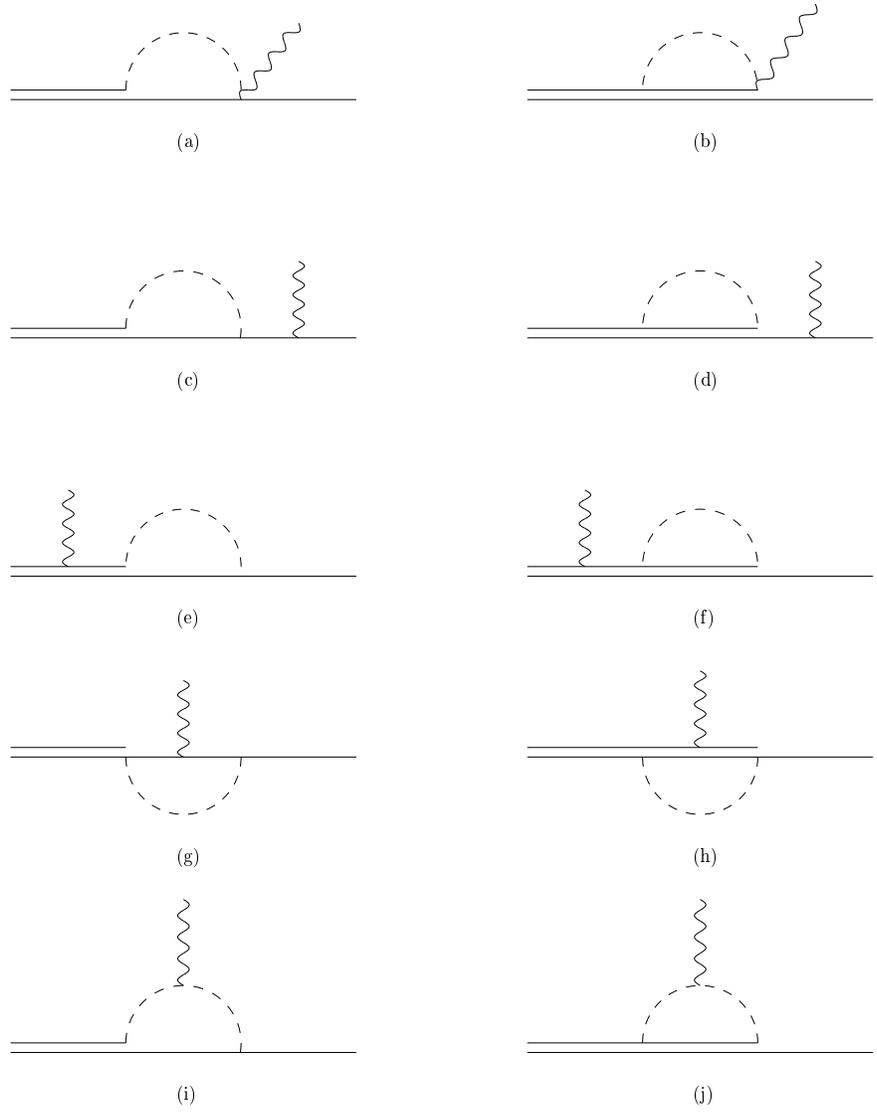,width=0.95\textwidth,angle=0}}
\caption[dummy]{Loop diagrams: single(double) solid lines denote
Nucleons(Deltas), respectively.}
\label{fig0}
\end{center}
\end{figure}

To ${\cal O}(\varepsilon^3)$ one can draw 10 independent loop-diagrams for
the $\Delta\rightarrow N\gamma$ transition (see Fig.~\ref{fig0}). However,
due to the constraints
$v\dt u_{v,\Delta}^i$ = $S\dt u_{v,\Delta}^i$  = $\tau_i u_{v,\Delta
\m}^i$ = 0 satisfied by the
on-shell spin 3/2 isospin 3/2 spinor $u_{v,\Delta \mu}^i$ only diagrams
1i, 1j in Fig. 1 survive. Their contribution can be formally written as
\begin{eqnarray}\label{HBXPT}
i{\cal M} =-\sqrt{\frac{2}{3}} {2 e g_{\pi N\Delta}\over F_\pi^2} \
\bar{u}_v(r_N)
\Bigl\{
  & (S\cdot \epsilon) q_\mu &
\left[ g_A F_N + \chi g_1 F_\Delta \right]
       \nonumber \\
+& (S\cdot q) \epsilon_\mu &
\left[ g_A G_N + \chi g_1 G_\Delta \right]
\nonumber \\
+& (S\cdot q) q_\mu &
\Bigl( \epsilon_0 \left[ g_A J_N + \chi g_1 J_\Delta \right]
      \Bigr.
\nonumber \\
&& + \Bigl.
     (q\cdot\epsilon) \left[ g_A H_N + \chi g_1 H_\Delta \right]
    \Bigr) \Bigr\} u_{v,\Delta}^{\mu}(0) \ .
\end{eqnarray}
Here the $N$ and $\Delta$ labels on the functions $F_i,G_i,...$ denote
the intermediate baryon propagator in
diagrams (i) and (j) of Fig.\ 1, respectively, while  the isospin factor
$\chi$ takes the value $ -{5\over 3}$. The loop functions are defined via
\begin{eqnarray}\label{functions}
F_N(t)&=& 2 \int_0^1  d x \ (x-1)
{A^{301}(\delta x,M^2_t)\over d} \nonumber \\
F_\Delta(t)  &=& 2 \int_0^1 dx \left[ (1-x) {A^{301}(-\delta
x,M^2_t)\over d}
              + 2x {A^{301}(-\delta x,M^2_t)\over d(d-1)}
 \right]\nonumber \\
G_N(t) &=& 2 \int_0^1  d x \ x
  {A^{301}(\delta x,M^2_t)\over d} \nonumber \\
G_\Delta(t) &=& -2 \int_0^1 dx \left[x {A^{301}(-\delta x,M^2_t)\over d}
              +2 (1-x) {A^{301}(-\delta x,M^2_t)\over d(d-1)}
\right]\nonumber \\
J_N(t) &=& \int_0^1  dx \ x(1-x)
A^{310}(\delta x,M^2_t)         \nonumber \\
J_\Delta(t) &=& \int_0^1  dx\ x(1-x) {d-3\over d-1}A^{310}(-\delta
x,M^2_t)
\nonumber \\
H_N(t) &=& \int_0^1  dx\ x(1-x) (1-2x) A^{300}(\delta x,M^2_t)
\nonumber \\
H_\Delta(t) &=& \int_0^1  dx\ x(1-x) (2x-1)
{d-3\over d-1}A^{300}(-\delta x,M^2_t)
 \ ,
\end{eqnarray}
where the $t\equiv q^2$ dependence arises implicitly through
$M^2_t=m_\pi^2+tx(x-1)$. Explicit representations for the $A^{ijk}(\W,\Mt)$ are
given in the Appendix. The $1/M^2$ sensitivity of our calculation is
saturated by the two inverse powers of the quantity $4\pi F_\pi\sim M$
which is of the order of the chiral symmetry breaking scale
$\Lambda_\chi(=\{4\pi F_\pi,M_N\})$.

We close the present subsection with a brief discussion concerning
renormalization issues. It is straightforward, using
Eq.(\ref{functions}) and
Eq.(\ref{exp1}), to see that Eq.(\ref{HBXPT}) for the loop
contributions to the decay amplitude of the $\Delta$(1232) is a {\em complex}
quantity
and contains several divergent pieces. The
lower-order couplings $b_1$, $b_6$ only possess a finite part, whereas the
infinities encountered in the loop amplitudes are absorbed by the infinite
parts of the ${\cal O}(\varepsilon^3)$
counterterms\footnote{The interaction terms
proportional to these two LECs are what one calls counterterms
in the conventional field theoretical language. Nevertheless, as is
usual in ChPT, we will sometimes employ the same name for the $b_1$, $b_6$
couplings.} $E_1$, $D_1$
\begin{eqnarray}
D_1&=&D_1^r(\lambda)+16 \pi^2 \beta_{D1} L \; , \nonumber \\
E_1&=&E_1^r(\lambda)+16 \pi^2 \beta_{E1} L\; ,
\end{eqnarray}
with \cite{BFHM}
\begin{equation}\label{divfactor}
L=\frac{\lambda^{d-4}}{16\pi^2}\left[\frac{1}{d-4}+\frac{1}{2}(\gamma_E-1-
\ln 4\pi)\right] .
\end{equation}
The two pertinent beta-functions
$\beta_{D1},\beta_{E1}$ can be found from Eq.(\ref{functions}) and the
expressions in Appendix A. It is straightforward to obtain the results
\begin{eqnarray}
\beta_{D_1}&=&\frac{g_{\pi N\Delta}M_N^2}{6\pi^2F_\pi^2}(g_A-5/9g_1)
\nonumber \\
\beta_{E_1}&=&\frac{g_{\pi N\Delta}M_N^2}{6\pi^2F_\pi^2}(g_A-20/9g_1).
\end{eqnarray}

At the moment, however, we have no information
about the numerical size of their finite parts
$E_1^r(\lambda),D_1^r(\lambda)$
at the chosen renormalization scale $\lambda$, as well as about the
magnitude of the coupling $b_6$. We note that this information will be
available in
the near future as several calculations regarding different scattering
processes in the delta resonance region using SSE are underway, which will
allow for a systematic extraction of the relevant higher order
$N\Delta$-couplings. In section VII we will nevertheless give some
numerical
estimates for linear combinations of these couplings based on calculations
for $N\Delta$-transition multipoles performed in theoretical frameworks
outside SSE.

\subsection{A note on gauge invariance}

Gauge invariance requires that, upon setting $\e^\m \rightarrow
q^\m$, the amplitude must vanish. The general amplitudes
Eq.(\ref{exp}) and Eq.(\ref{exp2}) satisfy this requirement explicitly
(recall
that $\e_0  \rightarrow q_0 = \delta $).
In the case of our loop calculation this requires
\begin{equation}\label{con}
 F_i(t) + G_i(t) + \delta J_i(t) +t H_i(t) = 0   \ ,
\end{equation}
separately for $i=N$ and $i=\D$.
This provides for a stringent test of our calculation. From the explicit
formulae
Eq.(\ref{functions}) and Eq.(\ref{exp1})
one can show that the constraints of Eq.(\ref{con}),
are indeed satisfied. This is most easily seen by partial
integration of the ${1\over\sqrt{\Omega^2-M^2_t}}$ part
of $(\delta J_i + t H_i)$.


\section{The $\Delta\rightarrow N\gamma^\ast$ form factors to
${\cal O}(\ve^3)$}

Combining the loop result, Eq.(\ref{HBXPT}), with the Born contributions,
Eq.(\ref{eq:born}),
 we obtain the total ${\cal O}(\ve^3)$ amplitude. This is to be
compared  with the general form of the amplitude, Eq.(\ref{exp2}). The
identification
of the three form factors is straightforward. At this point we have no
information about the magnitude of the counterterms $b_1,b_6
,D_1,E_1$. However,
the important observation is that the unknown counterterms only affect
the {\em overall normalization} of the 3 form factors $G_i(q^2)$ and leave
the {\em $q^2$-dependence unaffected}. We therefore separate the real
photon
point $q^2=0$ from each form factor and write
\begin{eqnarray}\label{loopG1}
G_1(q^2)=G_1(0)+\tilde{g}_1(q^2)  \nonumber \\
G_2(q^2)=G_2(0)+\tilde{g}_2(q^2)  \nonumber \\
G_3(q^2)=G_3(0)+\tilde{g}_3(q^2) \; ,
\end{eqnarray}
with $\tilde{g}_i(0)\equiv 0,\,i=1,2,3$.
It is then straightforward to arrive at
the results of Table I. Comparing the {\it
relativistic} form of the coupling structure proportional to $b_1$,
Eq.(\ref{rel1}), with
our general {\it relativistic} amplitude, Eq.(\ref{eq1neu}), one expects
that only
the $G_1$ form factor receives contributions from this coupling, which
also holds in the non-relativistic SSE formalism. Analogously, the
coupling $b_6$ in Table \ref{tab1} is shown only to contribute to
the form factor $G_2$, again as expected from its relativistic analogue in
Eq.(\ref{rel1}).
Furthermore, the results of Table \ref{tab1} are, as they should be,
gauge-independent. This happens automatically for the chiral counterterm
contributions. For the loop contributions, however,
gauge invariance can be explicitly demonstrated via the use of
Eq.(\ref{con}).

The results of Table \ref{tab1} can now be compared with the ones
in Ref.\cite{Lucio} where only the real-photon case is considered (no $G_3$).
We find a difference from our result originating in Eq.(\ref{exp}). In
particular,
there is no contribution to $G_2$ from the $b_1$
counterterm. This happens because, in \cite{Lucio}, the coefficient of
$G_1$ in the $(S\cdot q) (\epsilon\cdot  T_3)$ term is $2M_N$ rather than
$2M_N+\delta $.

In section \ref{numerics} we derive an estimate for the unknown couplings.
We will then be in position to describe both the $q^2$ dependence and
the
overall normalization of the three form factors.
Once again we stress
that {\em chiral symmetry demands that these form factors are complex
quantities},
which is due to
the on-shell $\pi N$ intermediate state of loop diagram (i) in Fig.1.

\section{Connection to $\Delta\rightarrow N\gamma^\ast$  Transition
Multipoles}\label{TrM}

In the relativistic case the identification of the transition
multipoles M1($q^2$),E2($q^2$),C2($q^2$)  
in terms of the  ``Dirac''-type form factors $G_1(q^2),G_2(q^2),G_3(q^2)$
is obtained \cite{definition} by reading off the spin-space tensor
properties
of the Pauli-reduced  most general amplitude, Eq.(\ref{eq1neu}),
in the ${\bf q}\cdot {\pmb \epsilon}=0$ gauge.
The explicit expressions read\footnote{At the photon point ($q^2=0$) one easily
verifies
that they agree with the corresponding expressions given in \cite{th}. We are
grateful
to Rick Davidson for valuable support in this point.}
\begin{eqnarray}
M1(q^2)&=&c_\Delta \left\{
G_1(q^2)[(3M_\Delta+M_N)(M_\Delta+M_N)-q^2] \right.\nonumber \\
& &\left. -G_2(q^2)\frac{M_\Delta}{2M_N}(M_\Delta^2-M_N^2-q^2)
-G_3(q^2)q^2\frac{M_\Delta}{\delta} \right\} \nonumber \\
E2(q^2)&=&-c_\Delta \left\{
G_1(q^2)(M_\Delta^2-M_N^2+q^2)-G_2(q^2)\frac{M_\Delta}{2M_N}(M_\Delta^2-M_N^2-q^2)
-G_3(q^2)q^2\frac{M_\Delta}{\delta}  \right\} \nonumber \\
C2(q^2)&=&-c_\Delta 2M_\Delta |\vec{q}|
\left\{G_1(q^2)-G_2(q^2)\frac{E_N}{2M_N}-G_3(q^2)\frac{\omega}{2\delta}\right\},
\end{eqnarray}
with
\begin{eqnarray}
c_\Delta=\frac{e|\vec{q}|}{12M_NM_\Delta\sqrt{2M_N(E_N+M_N)}}
\sqrt{\frac{2M_\Delta}{M_\Delta^2-M_N^2}}.
\end{eqnarray}
We now need to match these general expressions to the ${\cal O}(1/M^2)$
accuracy of
our results for the $G_1,G_2,G_3$ form factors obtained via SSE in the previous
section.
We may proceed in two ways:
\begin{enumerate}
\item Use the same reasoning as in the relativistic case,
      {\it i.e.}, obtain the multipoles
      from the spin tensor structure of the appropriate
      Pauli-reduced expression, Eq.(\ref{exp2}).
\item Expand the relativistic formulae
      for the multipoles to ${\cal O}(1/M^2)$.
\end{enumerate}
In both cases we obtain
\begin{eqnarray}\label{EMR}
 M1(q^2) & = &  {e\sqrt{\delta^2-q^2} \over 6  \sqrt{\delta}}\  \left[
                         \frac{2 G_1(q^2)}{M_N}
                      -  {\delta \over {4 M_N^2}} G_2(q^2)
                      -  {q^2 \over {4 \delta M_N^2}} G_3(q^2)+{\cal
                        O}(1/M_N^3)
                \right] \nonumber \\
 E2(q^2)  & = & {e\sqrt{\delta^2-q^2} \over 6 \sqrt{\delta}}\ \left[
                - {\delta\over 2 M^2_N}   G_1{(0)}
                + {\delta \over 4 M^2_N}    G_2(q^2)
                + {q^2 \over {4 \delta M^2_N}} G_3(q^2)+{\cal O}(1/M_N^3)
                 \right]
           \nonumber\\
 C2(q^2)   & = & {e(\delta^2-q^2) \over 6 \sqrt{\delta}} \
\left[
 -{1\over {2 M^2_N} }G_1{(0)}  + {1 \over {4 M^2_N} }G_2(q^2) +{1 \over {4
M^2_N}}
G_3(q^2)+{\cal O}(1/M_N^3)
\right] \ .
\end{eqnarray}
>From Eq.(\ref{EMR}) one can directly see that the ${\cal O}(\epsilon^3)$
prediction
for M1 is sensitive to subleading order, whereas the prediction for the
quadrupole
multipoles E2, C2 only gives us the leading result. One also observes the
peculiar
behavior that {\em in the case of the E2-, C2-form factors their
$q^2$-dependence
is not suppressed by powers of $1/M_N$ with respect to the $q^2=0$ point} !
According to
our knowledge this situation has not been observed\footnote{The notable
exception has been the pseudoscalar formfactor of the nucleon which is
known to be dominated by the light pion physics, \cite{BFHM}.}
 before in chiral
calculations
of form factors, e.g. compare \cite{BKM,BFHM}\footnote{The analogous phenomenon
occurs
for the slope parameters of the $G_i(q^2),\, i=1,2,3$ which we discuss in
section
\ref{predictG}.}. Its formal origin lies in the fact that
for quadrupole transitions
it is not possible to write down a counter term at a lower order than
${\cal O}(\epsilon^3)$, i.e. the same order where loops start contributing.
{\em
Keeping the above caveat in mind that Eq.(\ref{EMR}) only represents the
leading
behavior for the E2-, C2-form factors, the
chiral structure underlying Eq.(\ref{EMR}) leads us to the expectation that the
E2-,C2-form factors
only show a weak suppression at low $q^2$ with respect to their $q^2=0$ point
and that
the $q^2$-evolution could be very different from the corresponding (dipole
behavior of the) nucleon Sachs form
factors} $G_E(q^2), G_M(q^2)$, e.g. \cite{BKM,BFHM}. On the other hand, the M1
form
factor
shows the expected 1/M suppression of its radius with respect to the photon
point---quite
analogous to the SSE calculation of the dipole form factor $G_M(q^2)$ in
\cite{BFHM}.
This question will be
addressed in more detail in a future communication once precision information
on the
relevant coupling constants is available. We further note that the quadrupole
multipoles
in Eq.(\ref{EMR}) satisfy the long wavelength
($|{\bf q}|\rightarrow 0 $) constraint $E_2 =
{q_0\over |\bf q|} C_2$.


\section{Numerical Results}\label{numerics}

To provide numerical results we need to fix the
coupling constants entering the calculation. For our ${\cal O}(\epsilon^3)$
calculation it is appropriate to use
$F_\pi = 92.5$ MeV, $M_N=938$ MeV, $\delta=293$ MeV, $m_\pi=140$ MeV,
$g_A=1.26$, $g_{\pi N\D}=1.05$ \cite{BFHM,Compton2}. However, the four $\gamma
N\Delta$
couplings $b_1, b_6, D_1, E_1$ as well as the leading $\pi\Delta\Delta$
coupling $g_1$
are only poorly known at the moment. Until we have
more information on these couplings we therefore proceed {\em semi-empirically}
and
evaluate the {\em exact ${\cal O}(\epsilon^3)$ SSE results of sections II to
VI} with
phenomenological input
from other calculations.

In particular, for the leading M1 coupling we use
$b_1=7.7$ as in Ref.\cite{Compton2}\footnote{The factor of 2 difference
compared to
ref.\cite{Compton2} just results from the fact that we use a different form for
the ${\cal O}(\epsilon^2)$ $\gamma N\Delta$ vertex in Eq.(\ref{eq:a2}), in
order to be
consistent with the SSE formalism paper ref.\cite{long}. The sign of b1 is
chosen to
give a positive result for the ${\cal O}(\epsilon^2)$ M1 $\gamma N\Delta$
transition
multipole at $q^2=0$, consistent with the previous analyses \cite{th,disp}.},
although this is not entirely consistent with our
${\cal O}(\epsilon^3)$ calculation as this particular strength for the coupling
only holds to ${\cal O}(\epsilon^2)$ and undergoes an as yet undetermined
renormalization
in ${\cal O}(\epsilon^3)$. The remaining three $\gamma N\Delta$ couplings are
still
undetermined \cite{photo}, for now we find that we can parameterize them via
{\em 2 independent linear
combinations $C_2,C_3$} and fix them from Refs.\cite{disp,RPI} in the next
section.
Finally,
due to lack of more precise information at the moment, we use the SU(6) quark
model
estimate $g_1 = {9\over 5} g_A$ for the $\pi\Delta\Delta$ coupling in
Eq.(\ref{Lagrangian}), which should be accurate to 25\% ({\it e.g.}
\cite{couplings}).
We also note that the so called off-shell parameters $y_{1,6}$ of
Eq.(\ref{rel1})
decouple from our results.

\subsection{A best estimate for unknown couplings}

First, we note that the unknown couplings\footnote{For the numerical
analysis we utilize the scale-independent
couplings $\bar{D}_1$, $\bar{E}_1$.}
 $b_6,D_1,E_1$ only show up via two
independent linear combinations in our calculation. We therefore introduce the
new set of parameters
\begin{eqnarray}
C_2\equiv\frac{2\bar{E}_1-\bar{D}_1}{4}\,, \quad
C_3\equiv\frac{2b_6-\bar{D}_1}{4}\, .
\end{eqnarray}
We now want to utilize existing phenomenological analyses for $N\Delta$
transition
multipoles in order to fix these two parameters. Most of the existing work
relies on
parameterizations of pion photoproduction in the Delta region and has focused
on the
determination of the transition multipoles at the so called K-matrix pole
position,
e.g. see \cite{th}.
Unfortunately, this wealth of information is not directly transferable onto our
problem
because the K-matrix approach by construction only gives the imaginary part of
the
transition amplitude at the kinematical point where the real part vanishes.
For our calculation, however, we need information of the transition multipoles
at the
T-matrix pole, where $M1,E2,C2$ can be complex quantities even at the resonance
position.
This information has recently become available \cite{disp,VPI,RPI}. 
Our fitting procedure will refer to either the values of the real part of the transition
multipoles $M1,E2$ at $q^2=0$ or to the ratio E2/M1 at $q^2=0$. We shall
adopt the first
choice for the RPI,VPI results \cite{RPI} and the second one for the Mainz results
\cite{disp}.  
The two parameter sets determined by fitting
to ReM1(0), ReE2(0), with Re denoting the real part, of the RPI, VPI data 
will be designated
A and B respectively
while a further subscript taking the values 1 or 2 will denote the two fits used in 
reference \cite{RPI}. Specifically, using $ReM1(0)=0.3GeV^{-1/2}$,
$ReE2(0)=-0.0087GeV^{-1/2}$ we obtain
\begin{eqnarray}
C_2^{A_1}&=&-7.80 \nonumber \\
C_3^{A_1}&=&-2.48 ,
\end{eqnarray}
while using $ReM1(0)=0.301GeV^{-1/2}$, $ReE2(0)=-0.048GeV^{-1/2}$ we find 
\begin{eqnarray}
C_2^{A_2}&=&-8.07   \nonumber \\
C_3^{A_2}&=&-2.91.
\end{eqnarray}

We now consider the VPI group results. The values
$ReM1(0)=0.297GeV^{-1/2}$,
$ReE2(0)=-0.0065GeV^{-1/2}$ give
\begin{eqnarray}
C_2^{B_1}&=&-7.76     \nonumber \\
C_3^{B_1}&=&-2.72,
\end{eqnarray}
and finally the values $ReM1(0)=0.301GeV^{-1/2}$, $ReE2(0)=-0.0011GeV^{-1/2}$ produce
\begin{eqnarray}
C_2^{B_2}&=&-8.27   \nonumber \\
C_3^{B_2}&=&-3.31.
\end{eqnarray}
The next step is to use the speed-plot analysis\footnote{In view of the preliminary
character of these numerical estimates we neglect the fact that the Mainz numbers where
obtained at a pole position which would correspond to a smaller value for the
nucleon-delta mass-splitting
parameter $\delta$.} of the Mainz group \cite{disp}. This time we need our expression
\begin{equation}\label{my_emr}
\mbox{EMR}(0) \equiv \frac{E2(0)}{M1(0)}= { \delta\ b_1 + \delta G_2(0)
               \over
               8M_N G_1(0)  - \delta G_2(0) }
\end{equation}
from Eq.(\ref{EMR}) and table~\ref{tab1}.
We end up with a new parameter set, denoted by C
\begin{eqnarray}\label{fit}
\mbox{EMR}(0)_{Mainz} = (-3.5 - 4.6 \ i)\% &\rightarrow & C_2^{C}=-10.14
 \nonumber\\
                                           &\rightarrow & C_3^{C} =-2.25
\, .
\end{eqnarray}

All the fits produce ``natural size'' values for the parameters $C_2, C_3$. Based on the
above parameter sets we may determine an average value and a relevant error for C2, C3.
We arrive at the values
\begin{eqnarray}
C_2&=&-8.41\stackrel{\scriptstyle +}{-}0.44 \nonumber \\
C_3&=&-2.73\stackrel{\scriptstyle +}{-}0.18 .
\end{eqnarray} 

\subsection{Predictions for the Multipoles, CMR(0) ratio}

Having fixed $C_2,C_3$ in the previous subsection we can now give the
following
predictions for the average values of the electromagnetic form factors and their
relevant errors
\begin{eqnarray}
M1(0)&=& (308.6\stackrel{\scriptstyle +}{-}8.9 + \ i 28.6 )10^{-3}GeV^{-1/2} \nonumber
\\
E2(0)&=& (-6.4\stackrel{\scriptstyle +}{-}1.7 - \ i 16.9)10^{-3}GeV^{-1/2} \nonumber
\\
C2(0)&=& (-10.8\stackrel{\scriptstyle +}{-}1.7 - \ i 14.3)10^{-3}GeV^{-1/2} .
\end{eqnarray}
For the ratios we determine
\begin{eqnarray}
ReEMR(0)\equiv Re\frac{E2(0)}{M1(0)}&=(-2.52\stackrel{\scriptstyle +}{-}0.48)\% , \\
& & \nonumber \\
ImEMR(0)\equiv Im\frac{E2(0)}{M1(0)}&=(-5.24\stackrel{\scriptstyle +}{-}0.16)\% , \\
& & \nonumber \\
ReCMR(0)\equiv Re\frac{C2(0)}{M1(0)}&=(-3.86\stackrel{\scriptstyle +}{-}0.47)\% , \\
& & \nonumber \\
ImCMR(0)\equiv Im\frac{C2(0)}{M1(0)}&=(-4.29\stackrel{\scriptstyle +}{-}0.14)\% .
\end{eqnarray}

We note that the imaginary
parts of the multipole form factors (as well as the imaginary parts of
the form factors $G_i$ $i\in\{1,2,3\}$) are the same for all fits and so we are not
in a position to produce error bars for them. The
reason is simply that the imaginary part of the transition amplitude stems
exclusively
from loop contributions which do not depend on $C_2,\, C_3$.
 Of course the same is not true for the
ratios of the multipoles since in this case real and imaginary parts of
the numerator and the denominator are mixed in order to obtain the
real and imaginary part of the ratio. 
We also note that our results appear much less sensitive to $C_2$ compared with their 
dependence on $C_3$. The reason is that this first parameter
appears only in the $G_1$ form factor and its contibution with respect to
the other ones is suppressed by a power of the ratio $\frac{\delta}{M_N}$.

\begin{figure}[htb]
\begin{center}
\mbox{\epsfig{file=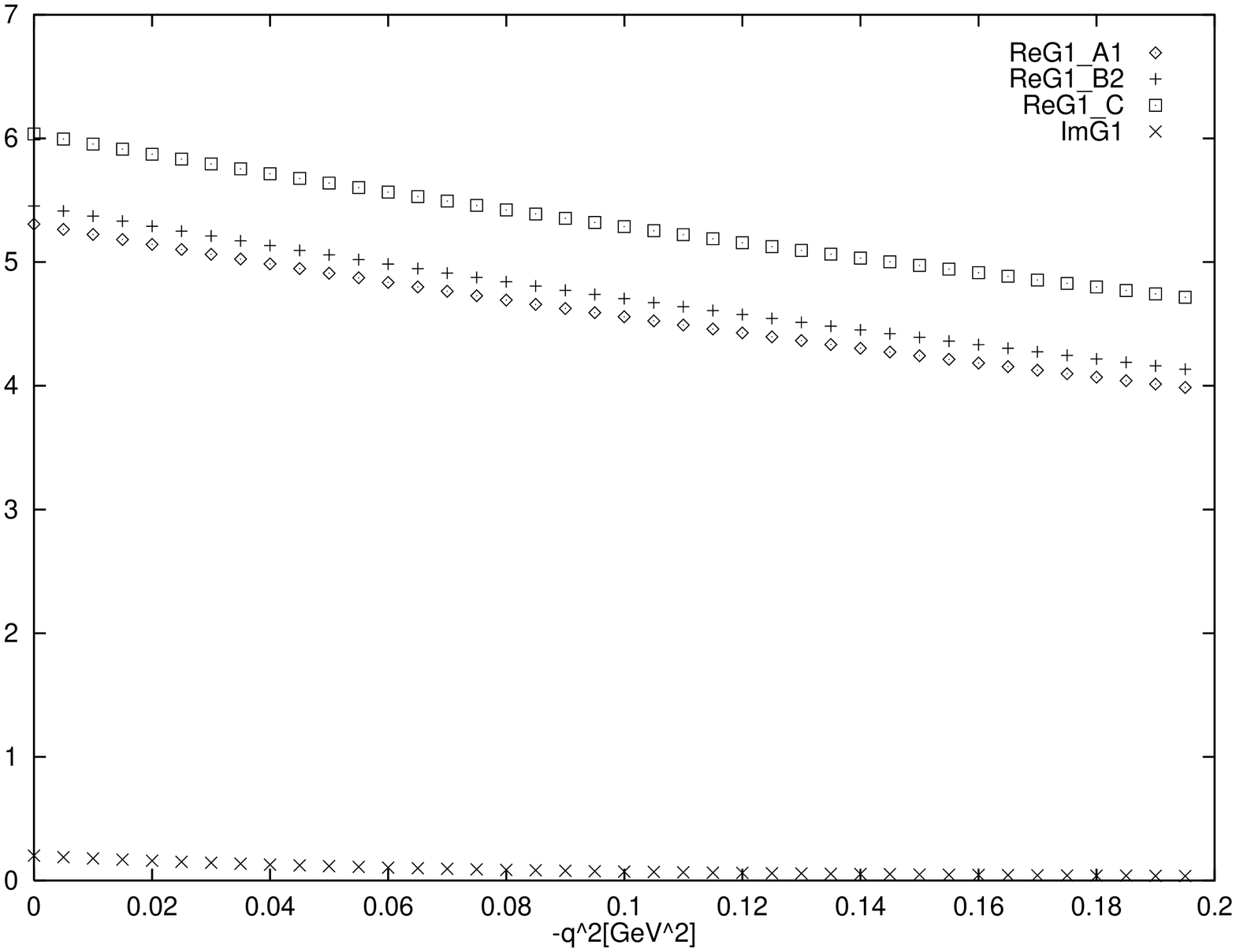,width=0.8\textwidth,angle=-90}}
\caption[dummy]{The real and the imaginary part of the form factor G1 with
$C_2$, $C_3$ given by the sets $A_1$, $B_2$, $C$.}
\label{fig3}
\end{center}
\end{figure}

\begin{figure}[htb]
\begin{center}
\mbox{\epsfig{file=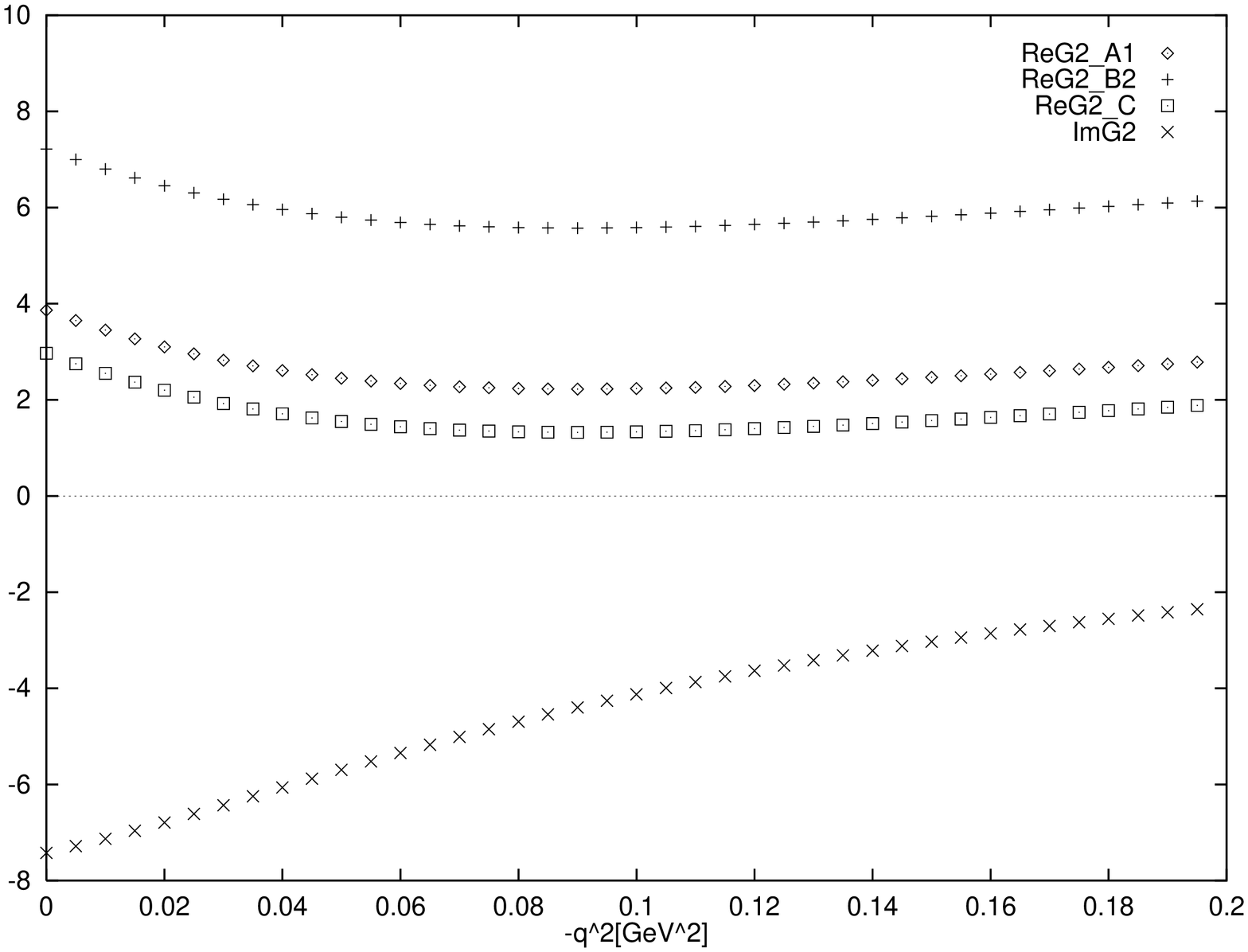,width=0.8\textwidth,angle=-90}}
\caption[dummy]{The real and the imaginary part of the form factor G2
with $C_2$, $C_3$ given by the sets $A_1$, $B_2$, $C$.}
\label{fig4}
\end{center}
\end{figure}

\begin{figure}[htb]
\begin{center}
\mbox{\epsfig{file=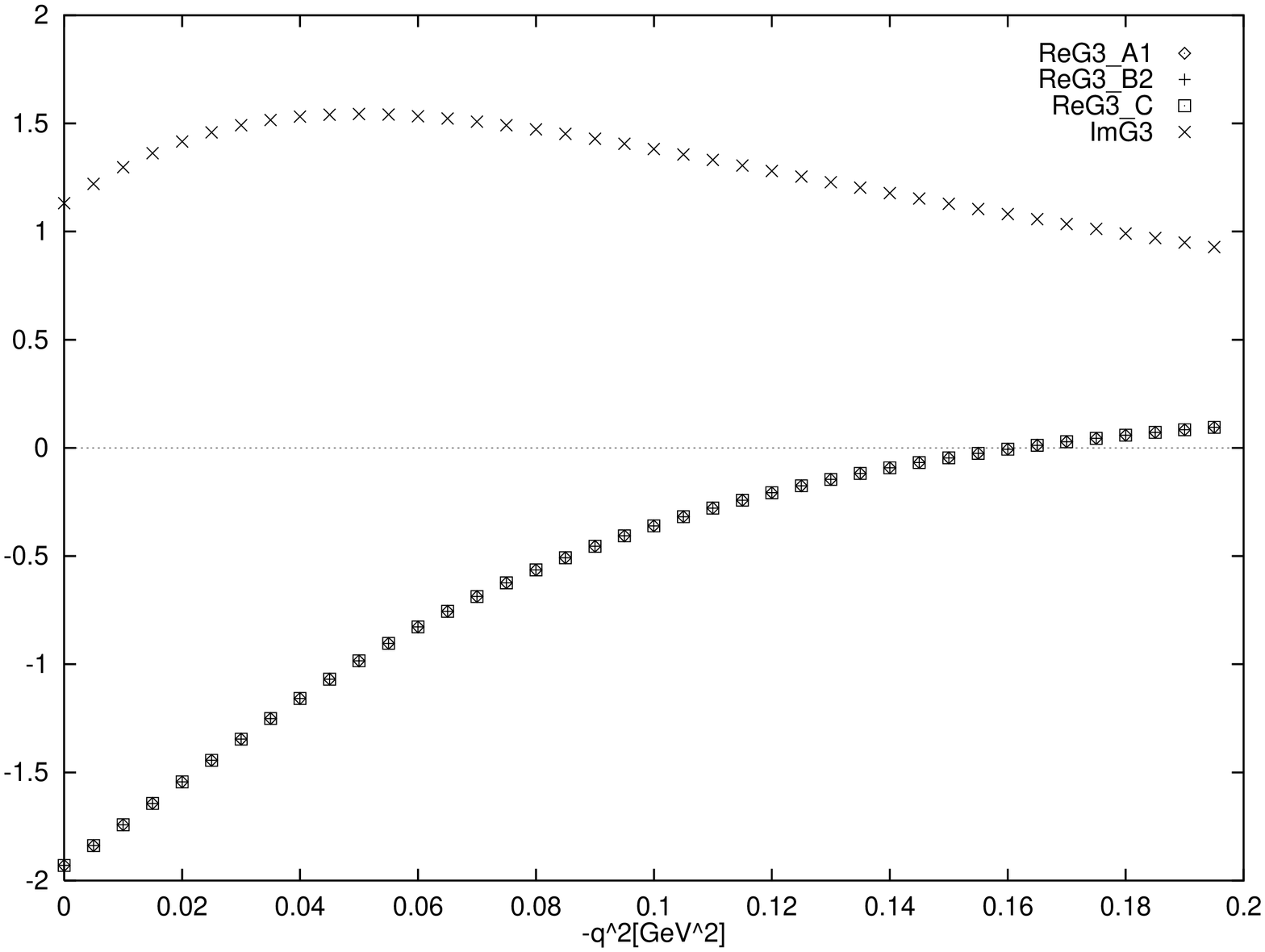,width=0.8\textwidth,angle=-90}}
\caption[dummy]{The real and the imaginary part of the form factor G3
with $C_2$, $C_3$ given by the sets $A_1$, $B_2$, $C$.}
\label{fig5}
\end{center}
\end{figure}

\begin{figure}[htb]
\begin{center}
\mbox{\epsfig{file=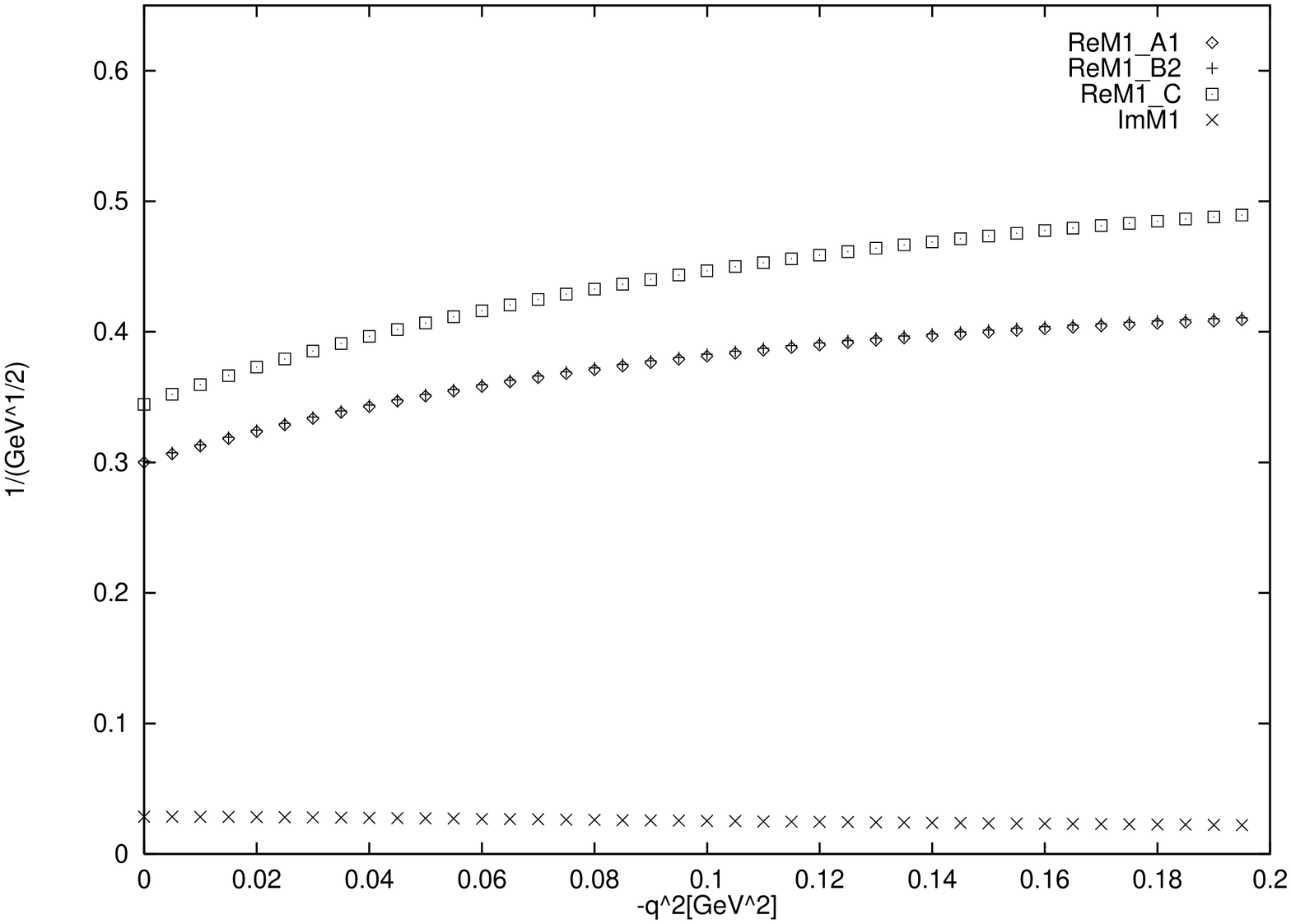,width=0.8\textwidth,angle=-90}}
\caption[dummy]{The real and the imaginary part of the form factor M1
with $C_2$, $C_3$ given by the sets $A_1$, $B_2$, $C$.}
\label{fig6}
\end{center}
\end{figure}

\begin{figure}[htb]
\begin{center}
\mbox{\epsfig{file=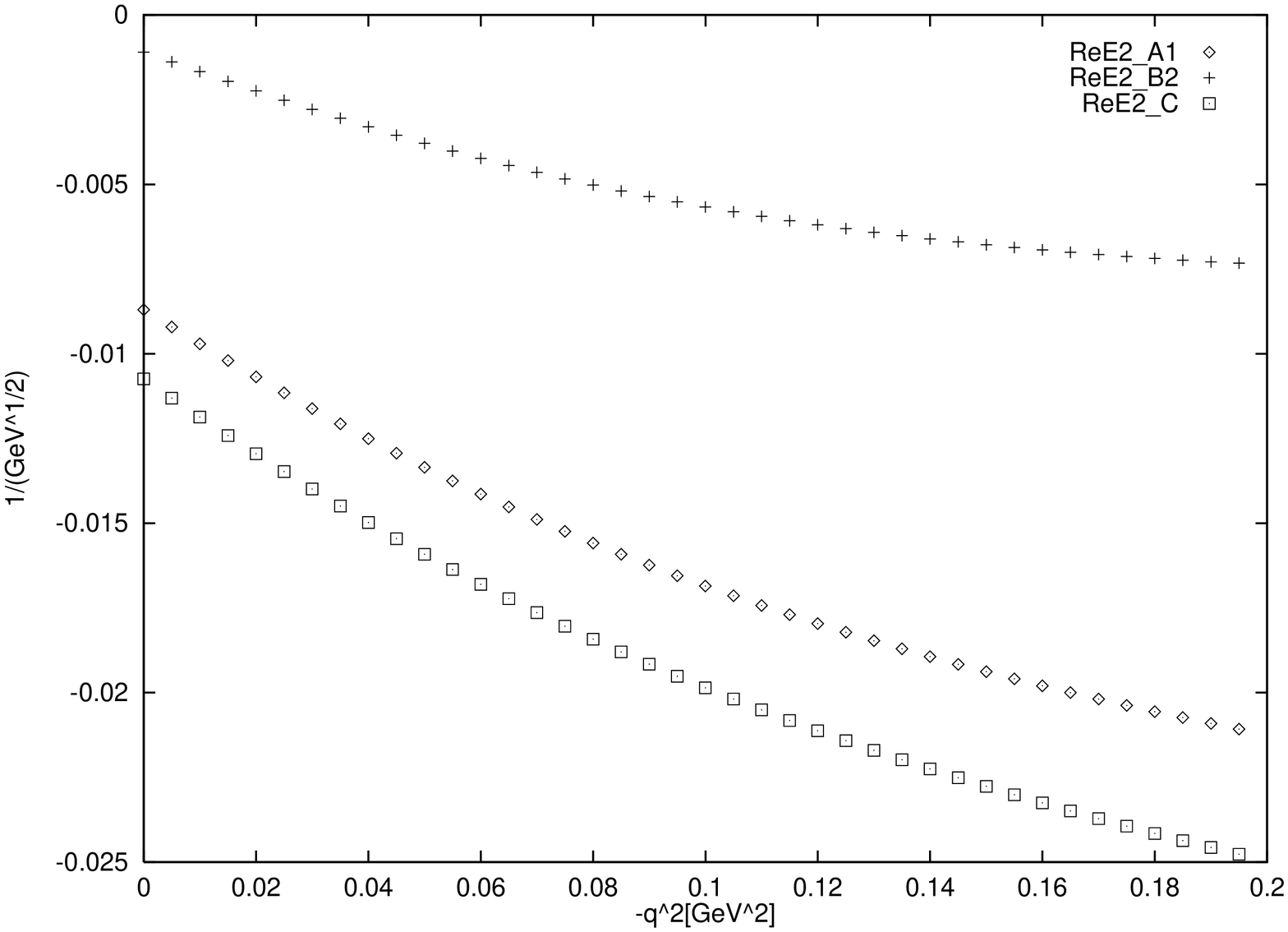,width=0.8\textwidth,angle=-90}}
\caption[dummy]{The real part of the form factor E2
with $C_2$, $C_3$ given by the sets $A_1$, $B_2$, $C$.}
\label{fig7}
\end{center}
\end{figure}

\begin{figure}[htb]
\begin{center}
\mbox{\epsfig{file=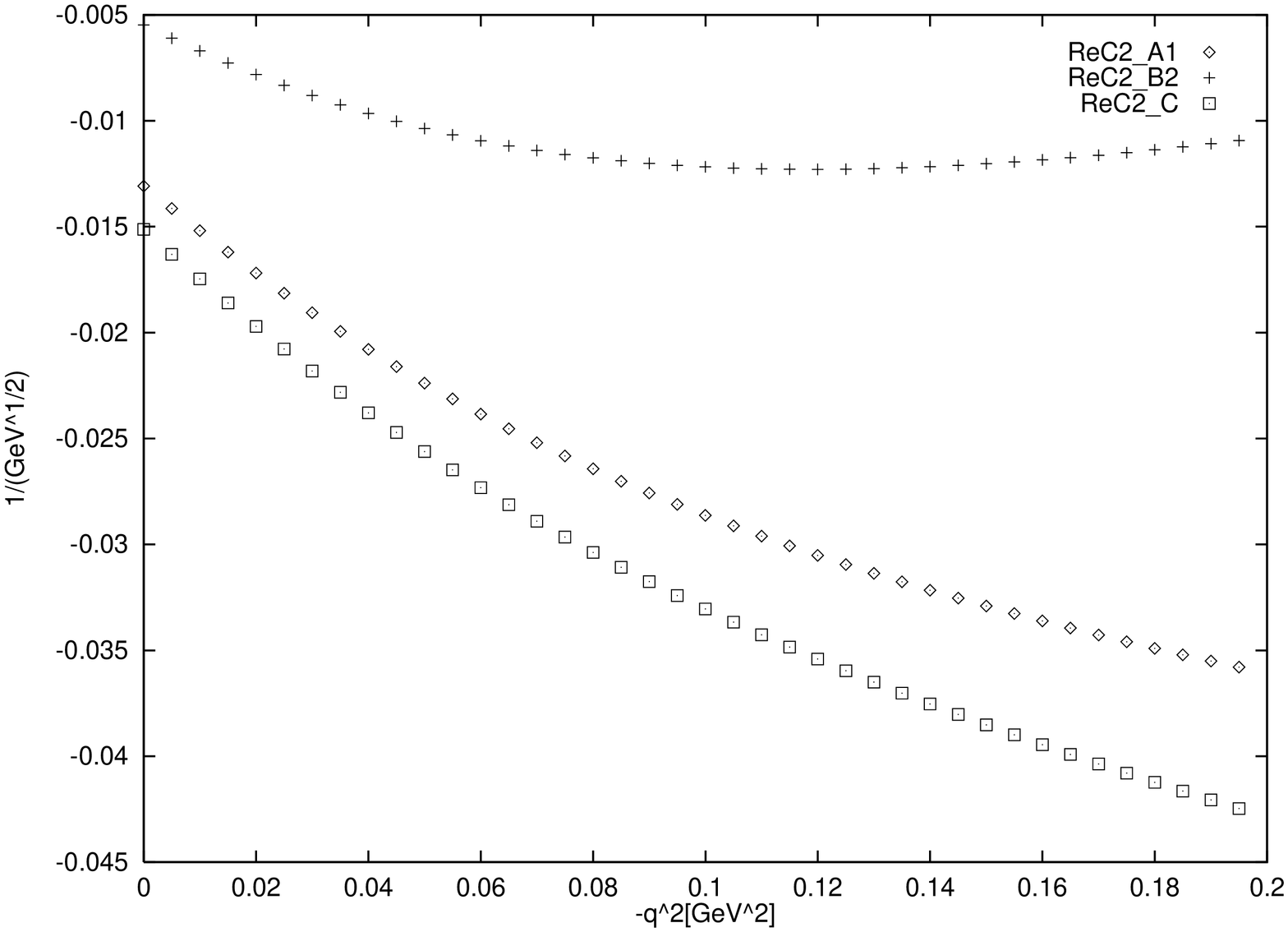,width=0.8\textwidth,angle=-90}}
\caption[dummy]{The real part of the form factor C2
with $C_2$, $C_3$ given by the sets $A_1$, $B_2$, $C$.}
\label{fig8}
\end{center}
\end{figure}

\begin{figure}[htb]
\begin{center}
\mbox{\epsfig{file=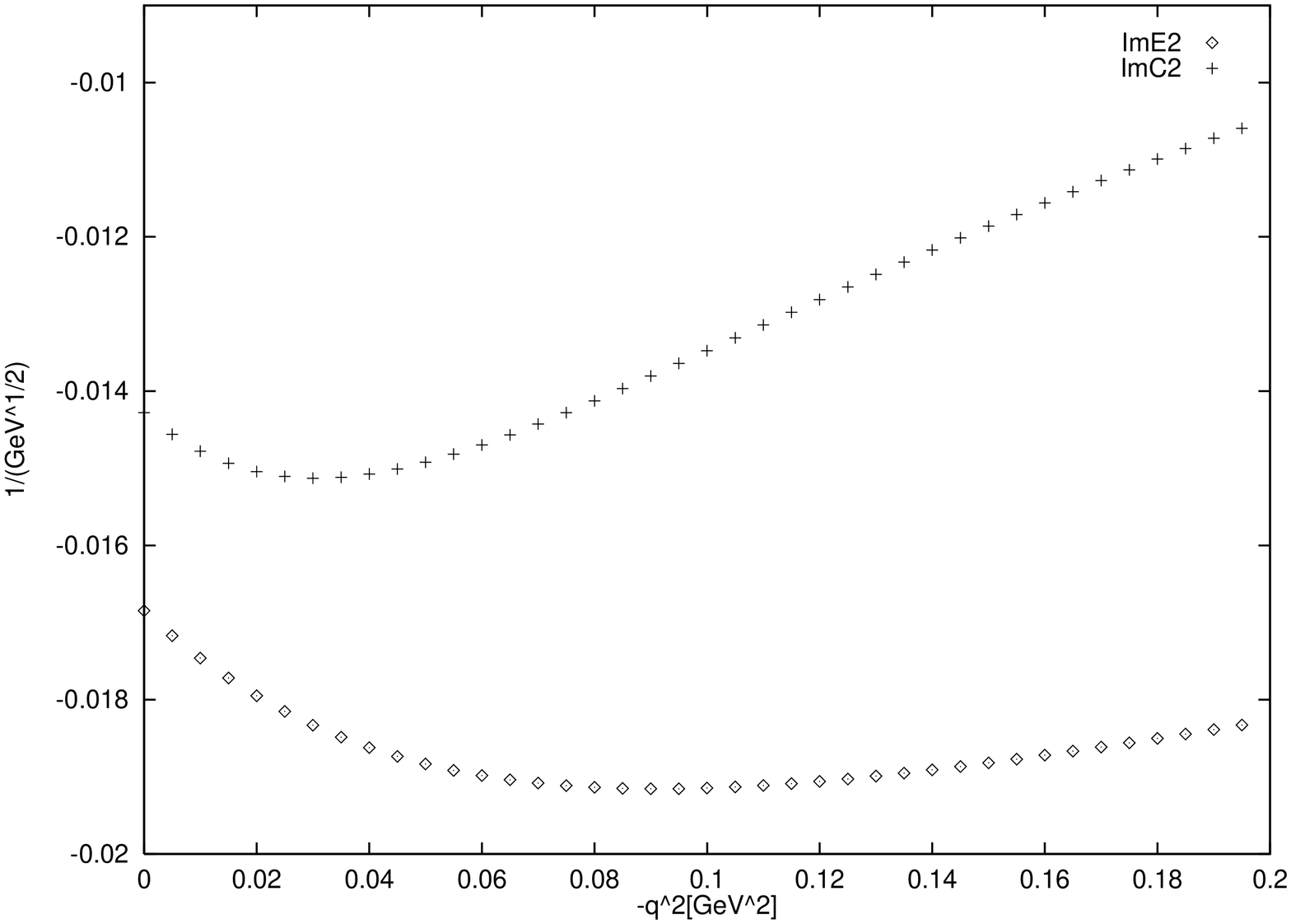,width=0.8\textwidth,angle=-90}}
\caption[dummy]{The imaginary part of the form factors E2, C2
with $C_2$, $C_3$ given by the sets $A_1$, $B_2$, $C$.}
\label{new}
\end{center}
\end{figure}

In Figures~\ref{fig3}, \ref{fig4}, \ref{fig5} the $q^2-$dependence of both
the real and the imaginary parts of $G_1$, $G_2$, $G_3$ is depicted,
using the C, $A_1$ and $B_2$ sets, respectively, for the values of the
parameters $C_2$, $C_3$. 
As for the real part of the form factor $G_3$ we do not want to imply that it
changes its sign near $-q^2=0.15 GeV^2$. Given that $G_3(q^2)$ approaches quickly to
zero even small corrections from higher orders can change the result for $-q^2>0.1$
$GeV^2$.
The resulting $q^2-$dependence of the multipole form
factors $M1$, $E2$, $C2$ is shown in Figures~\ref{fig6}, \ref{fig7},
\ref{fig8} and \ref{new} respectively.
Finally the last remaining figures show the real and the imaginary
parts of the ratios EMR, CMR. We close the present section with a remark
concerning the range of $q^2$, for which we believe our results to be
meaningful.
As we have already
mentioned in the previous paragraph, the {\em imaginary} part of the
amplitudes is determined entirely by a 1-loop diagram related to $\pi
N$-scattering
at the relatively large kinematic energy of approximately $\delta-m_\pi\sim
150$ MeV.
We therefore do not necessarily expect that
the ${\cal O}(\epsilon^3)$ calculation leads to a
good description of the {\em imaginary parts} and point to the {\em
possibility} of
large higher order corrections in the {\em imaginary parts} of the $G_i(q^2)$
form factors! This issue can only be resolved by a future
${\cal O}(\epsilon^4)$ calculation.

\subsection{Predictions for the Form Factors}\label{predictG}

Though enjoying less fame than the corresponding transition multipoles of the
previous
section, the three $N\Delta$ transition form factors $G_i(q^2)$ are actually
better
suited for a chiral analysis---the reason being that to ${\cal O}(\epsilon^3)$
the
complete $q^2$-dependence is solely governed by ${\cal O}(\epsilon)$
parameters, whereas
the poorly known couplings $b_1,b_6,D_1,E_1$ only enter in the $q^2=0$
normalization
of the form factors.
Utilizing $C_2,C_3$ from the previous section
one fixes
\begin{eqnarray}\label{Gpredict}
G_1(0)&=& (5.50\stackrel{\scriptstyle +}{-}0.14 + \ i 0.20)  \nonumber \\
G_2(0)&=& (4.90\stackrel{\scriptstyle +}{-}0.73 - \ i 7.42)  \nonumber \\
G_3(0)&=& (-1.93 + \ i 1.13)  .
\end{eqnarray}
We note that all
three
imaginary parts
and the real part of $G_3(0)$ do not depend on the couplings
$b_1,b_6,D_1,E_1$
and are
given by the much better known leading order parameters of SSE.

We now proceed to give predictions for the slopes parameters\footnote{We
focus on the
slope
parameters rather than on the transition radii in order to avoid the appearance
of any
of the poorly known couplings $b_1,b_6,D_1,E_1$ in the discussion.}
\begin{equation}\label{rad_def}
\rho^2_i \equiv 6 {d G_i \over dt}|_{t=0} \, ,
\end{equation}
which can be directly calculated from Eqs.(42-44).
Defining
$\m\equiv{m_\pi\over\delta}$ and ${\cal N}$ = $-{3 g_{\pi N\Delta} g_A
M_N
\over 4 \pi^2 F_\pi^2 \delta}$ we find
\begin{eqnarray}\label{radii}
\rho^2_1 &=& {\cal N}\
            \biggl[ -{3\over 2} + 2\,\m\,\pi  -
  {i\over 3}\,\sqrt{-1 + \m^{-2}}\,\m\,\pi +
  {2\over 3}\,\m^3\,\pi
\nonumber\\&-&
  {5\over 8} \,\m^2\,\pi^2
 + {\m\over 3} \sqrt{-1 + \m^{-2}}
      \log (-\sqrt{-1 + \m^{-2}} + {1\over \m})
\nonumber\\&+&
  {2\over 3}\,\m^3\,\sqrt{-1 + \m^{-2}}\,
      \log (-\sqrt{-1 + \m^{-2}} + {1\over \m})
\nonumber\\&+&
  {\m^2\over 2}\,\log^2(-\sqrt{-1 + \m^{-2}} + {1\over \m})
+ {4\over 3}\,\m\,\sqrt{-1 + \m^{-2}}
   \log (\sqrt{-1 + \m^{-2}} + {1\over \m})
\nonumber\\&+&
  {2\over 3}\,\sqrt{-1 + \m^{-2}}\,\m^3\,
      \log (\sqrt{-1 + \m^{-2}} + {1\over \m})
  -{2\over 3}\,i\,\sqrt{-1 + \m^{-2}}\,\m^3\,\pi
\nonumber\\&+&
  i\, \m^2\,\pi \,\log (\sqrt{-1 + \m^{-2}} + {1\over \m}) -
  \m^2\,\log^2(\sqrt{-1 + \m^{-2}} + {1\over \m}) \biggr] \ .
\end{eqnarray}
One can see immediately that $\rho_1^2$ is suppressed by a power of
$1/\Lambda_\chi\approx1/M_N$ compared to the real photon point $q^2=0$. This
behavior
can also be observed
in the $F_2(q^2)$ Dirac form factor of the nucleon (e.g. \cite{BKM,BFHM}). An
analogous
result is expected for the radius of the M1($q^2$) form factor, see the
discussion in
section \ref{TrM}.

For the slope parameters corresponding to $G_2$ and $G_3$, $t$-differentiation
of
the $J$, $H$ functions is needed, resulting in two divergent
contributions (one non-integrable singularity resulting
from differentiating under the integral sign and one divergent
surface term from differentiating w.r.t. the integration limit).
Treating the integral as a principal-valued one the infinities
are shown to cancel out, leaving
\begin{eqnarray}\label{radii2}
\rho^2_2 &=&
           4\pi{\cal N}\frac{M_N}{\delta} \left[  2\mu -{3\over
2}\pi\mu^2
+ {8\over 3}\mu^3
      - i\Bigl[ \sqrt{1-\mu^2} \left(1+8\mu^2 \over 3\right)
              - 3\mu^2\log( {1\over \m}+\sqrt{{1\over \m^2}-1})
          \Bigr] \right] \\
\rho^2_3 &=&  -2 \pi {\cal N}\frac{M_N}{\delta} \left[
               8\m +{16\over 3}\m^3 - (1+{15\over 2}\m^2) \left(
                  {\pi\over 2} - i\log({1\over\m} +
                \sqrt{{1\over\m^2}-1}\right)
         -i\sqrt{1-\m^2}\left({19\over 6} + {16\over 3}\m^2\right)
      \right] \nonumber .
\end{eqnarray}
It is important to note that this is a quite peculiar result. One can see
immediately
that these two slope parameters\footnote{The same of course holds for the
corresponding
radii.} scale as $\Lambda_\chi^{\, 0}\approx M_N^{\, 0}$, i.e. they show no
suppression
by a heavy scale with respect to the $q^2=0$ point! This behavior is completely
analogous to the one expected for $E2(q^2), C2(q^2)$ (cf. section \ref{TrM}),
but has
not yet been observed in chiral calculations of baryon form factors. The first
moment
with respect to $q^2$ of these ``quadrupole-like'' form factors
$G_2(q^2),G_3(q^2)$ is
entirely given by the small scales $m_\pi,\delta$ and leading order coupling
constants---allowing, {\em in principle}, a large
variation of the form factors at low $q^2$ compared to their $q^2=0$ point.
However,
we want to emphasize that Eqs.(\ref{radii},\ref{radii2}) only represent the
leading
results for the three slope parameters, more cannot be learned from a
${\cal O}(\epsilon^3)$ calculation. At present, we have no way of knowing the
size of
the ${\cal O}(\epsilon^4)$ corrections, but in principle it is possible to
calculate
them at a later stage\footnote{At first one would need a precise determination
of $b_1$
and the subleading corrections to the $\pi N\Delta$ vertex, as these couplings
would
enter the new loop diagrams of ${\cal O}(\epsilon^4)$.}.

Finally, we can obtain the chiral limit ($\m\rightarrow 0$) behavior of the
slope
parameters from
Eqs.(\ref{radii}, \ref{radii2}),
\begin{eqnarray}\label{r1chiral}
\rho^2_1 & = & {\cal N} \left[-{3\over 2} - {{i\,\pi }\over 3} +
    \log 2 - \log \m
              \right] +  {{\rm O}(\m)}                       \nonumber\\
\rho^2_2 & = & -{4\over 3} i\pi{\cal N}\frac{M_N}{\delta}  +
                {{\rm O}(\m)}   \nonumber\\
\rho^2_3 & = & - 2\pi {\cal N} \frac{M_N}{\delta} \left[ -{\pi\over 2} +
               i( \log 2 - \log\m - {19\over 6} )
                    \right]  + {\rm O}(\m)\ .
\end{eqnarray}
Thus, the real part of $\rho^2_1$ diverges logarithmically
in the chiral limit while its imaginary part remains finite.
On the other hand,  the real  part of  $\rho^2_2$
vanishes altogether in the chiral limit, while its imaginary part remains
finite. As for $\rho^2_3$, it is the real part that remains finite in the
chiral limit while its imaginary part diverges logarithmically.
The numerical values for the slope parameters are also collected in
Table~\ref{tab2} and since they take contributions only from the loops are 
independent from the values of $C_2$, $C_3$.
\section{Summary and Outlook}

The main features of our investigation
can be summarized as follows:
\begin{itemize}
\item
      The $\Delta\rightarrow N\gamma^\ast$ amplitude has been calculated to
      ${\cal O}(\epsilon^3)$ in the ``small scale expansion'' formalism.
      The pertinent counterterm contributions, relevant 1/M corrections to
lower
      order coupling and all loop diagrams allowed to this order have been
      analyzed systematically. It was found that the $q^2$-dependence of the
transition
      is generated by two $\pi N-,\pi\Delta -$loop diagrams and that two
counterterms
      are needed to renormalize the result.
\item
     We have obtained an estimate for presently poorly known couplings from
existing
     phenomenological analyses and then given 
predictions for the complex transition multipoles $M1(q^2)$,
$E2(q^2)$, $C2(q^2)$ as well as the complex
multipole
     ratios CMR(0), EMR($q^2$), CMR($q^2$). While matching the
     ${\cal O}(\epsilon^3)$ results to the corresponding expressions of the
multipoles
     as functions of $q^2$ we found that the radii of the quadrupole form
factors
     are not suppressed by a large scale compared to the $q^2=0$ point,
indicating
     a possibly rapid variation at low $q^2$. Further study of this interesting
result
     will continue once more precise information on the necessary couplings is
available.
\item
      Three appropriately defined electromagnetic $N\Delta$ transition form
factors
      $G_i(q^2),\,i=1,2,3$
      have been identified from the ${\cal O}(\epsilon^3)$
      $\Delta\rightarrow N\gamma^\ast$ calculation. The longitudinal form
factor
      $G_3(q^2)$ had to be rescaled compared to existing definitions in the
literature
      in order to achieve consistency with the power-counting of SSE.
      $G_1(q^2)$ could be found to
      subleading order, whereas for $G_2(q^2),G_3(q^2)$ we
      obtained the leading result.
\item
      The three $G_i(q^2)$ form factors are found to be complex quantities due
to
      the $\pi N$-loop
      diagram (i) in Fig.1. Their entire $q^2$dependence is controlled
      by relatively well-known ${\cal O}(\epsilon)$ parameters---making the
$G_i(q^2)$
      form factors the prefered testing ground of chiral symmetry in the
radiative
      $N\Delta$-transition. The corresponding slope
      parameters have been calculated and their chiral limit behavior has been
      discussed.
\end{itemize}
The numerical results of this work have to be considered preliminary due to the
presently
poor state of knowledge on coupling constants in the SSE formalism. However,
this
situation is going to improve in the near future and we will at a later point
revisit
our ${\cal O}(\epsilon^3)$ analytical results and try to improve upon the
numerical
accuracy of our predictions.

\begin{center}
{\bf Acknowledgments}
\end{center}
We would like to thank  V. Bernard, R. Davidson, U.-G. Mei{\ss}ner, M. Neubert,
C.N.
Papanicolas and R. Workman for valuable  discussions. C. N. K. also wishes to
acknowledge
several interesting exchanges with A. Bernstein.

The research of C. N. K. and G. C. G. has been supported in part by the
EU Programme ``Training and Mobility of Researchers", Network
``Hadronic Physics with High Energy
Electromagnetic probes", contract ERB FMRX-CT96-0008. Finally, the research
of G.P. has been financed under ERBFMBICT 961491 Return Marie Curie
Fellowship. T. R. H. would like to thank the National Institute for
Nuclear Theory
(INT) in
Seattle for its hospitality, where part of this work was done.

\appendix
\section{Integrals}
The $A^{ijk}(\W,\Mt)$ functions used in Eq.(\ref{HBXPT}) are defined as
Euclidean integrals in dimensional regularization
\begin{equation}\label{defa}
A^{ijk}(\W,\Mt) \equiv \int_0^\infty \a^j \ d\a \int
{d^d l_E\over (2\pi)^d} {(-l^2_E){}^k\over (-l^2_E -{1\over 4} \a^2
    + \a \W - \Mt )^i}  \ .
\end{equation}
        To calculate them, first of all, we express the denominator
using the Schwinger proper time representation
\begin{equation}
\int^{\infty}_{0}dxx^{n-1}e^{-Ax}=\frac{\Gamma(n)}{A^{n}} \hspace{0.1in}
A>0
\end{equation}
with $\Gamma(n)$ the well-known $\Gamma$ function. The constraint $A>0$
is explicitly fulfilled for the $\Omega$ $<0$ case while for $\Omega$ $>0$
we analytically continue our expressions. Finally using the tabulated
integrals
\begin{equation}
\int_{0}^{\infty}x^{\nu-1}e^{-\beta x^2-\gamma x}dx=(2\beta)^{-(\nu/2)}
\Gamma(\nu)\exp(\frac{\gamma^2}{8\beta})D_{-\nu}\left(\frac{\gamma}
{(2\beta)^{1/2}}\right)
\end{equation}
\begin{equation}
\int_{0}^{\infty}e^{-zt}t^{-1+\frac{\beta}{2}}D_{-\nu}[2(kt)^
{\frac{1}{2}}]dt=\frac{2^{1-\beta-\frac{\nu}{2}}\Gamma(\beta)}
{\Gamma(\frac{1}{2}\nu+\frac{1}{2}\beta+\frac{1}{2})}(z+k)^
{-\frac{\beta}{2}}F\left(\frac{\nu}{2},\frac{\beta}{2};\frac
{\nu+\beta+1}{2};\frac{z-k}{z+k}\right)
\end{equation}
with $D_{-\nu}(x)$ the parabolic cylinder function and
F(a,b;c;x) the hypergeometric function,
we find explicitly
\begin{eqnarray}\label{exp1}
{A^{301}\over d} &=&
-L\W + {\W\over 32 \pi^2}\left(1-\ln{\Mt\over\mu^2}\right)
              +{1\over 16\pi^2} \times
 \left\{ \begin{array}{l}
          -\smw\aco \\
          +\swm\loga \\
 \end{array} \right. \nonumber \\
{A^{301}\over d(d-1)} &=&
-{1\over 3}L\W + {\W\over 288 \pi^2}\left(5 - 3\ln{\Mt\over\mu^2}\right)
              +{1\over 48\pi^2} \times
 \left\{ \begin{array}{l}
          -\smw\aco \\
          +\swm\loga \\
 \end{array} \right. \nonumber \\
 A^{310} &=&
  2L +{1\over 16 \pi^2} \left(1+ \ln{\Mt\over\mu^2}\right)
-{\W\over 8\pi^2}\times
 \left\{ \begin{array}{l}
          {1\over \smw}\aco \\
          {1\over\swm}\loga \\
 \end{array} \right. \nonumber \\
{d-3\over d-1} A^{310} &=&
{2\over 3}L +{1\over 144 \pi^2} \left(7+ 3\ln{\Mt\over\mu^2}\right)
-{\W\over 24\pi^2}\times
 \left\{ \begin{array}{l}
          {1\over \smw}\aco \\
          {1\over\swm}\loga \\
 \end{array} \right. \nonumber \\
A^{300} &=&
-{1\over 16\pi^2}\times
\left\{ \begin{array}{l}
{1\over \smw}\aco \\
          {1\over\swm}\loga \\
 \end{array} \right.  \nonumber \\
{d-3\over d-1} A^{300} &=&{1\over 3} A^{300}
\ ,
\end{eqnarray}
where the factor L
carries the infinity in
dimensional regularization and is defined in Eq.
(\ref{divfactor}). It is clear that
$A^{300}$ is finite while  $A^{310}$ and $A^{301}$ diverge.
The integrals in Eq.(\ref{functions}) are split between $[0,x_0]$
(trigonometric branch)
and $(x_0,1]$ (log branch) where $x_0$ is the large root of $\W^2=\Mt$,
$$x_0 = {-t+\sqrt{t^2+4m_\pi^2|{\bf q}|^2} \over 2|{\bf q}|^2 } \ ,$$
with $t=\delta^2-|{\bf q}|^2$ to the order we work.
In the calculation the functions are defined with
$\W = \delta x$ for the $N\pi$ intermediate state loop diagram
and $\W = -\delta x$ for the  $\Delta\pi$ one. This entails
(cf. Eq.(\ref{exp1})) that the logarithms have negative arguments
(and therefore there is an absorptive piece of the amplitude)
for the $N\pi$ diagram alone, as expected from general considerations
($s=M_\Delta^2 \ge (M_i + m_\pi)^2$ for $i=N$ only). The imaginary
parts are computed analytically.

For completeness, we show the correspondence between our $A$ functions
and the $J$ functions appearing in the literature (e.g., ~\cite{Compton2}):
\begin{eqnarray}
A^{200}(\W,\Mt)   &=& J_0(\W,\Mt)   \nonumber \\
A^{201}(\W,\Mt)   &=& d J_2(\W,\Mt) \nonumber \\
A^{300}(\W,\Mt)   &=& {1\over 2} {\partial\over \partial_{\Mt}}
                                 J_0(\W,\Mt) \nonumber \\
A^{301}(\W,\Mt)   &=& {d\over 2} {\partial\over \partial_{\Mt}}
                                 J_2(\W,\Mt) \nonumber \\
A^{310}(\W,\Mt)   &=& -{1\over 2} {\partial\over \partial_{\W}}
                                 J_2(\W,\Mt) =
                      {\partial\over \partial_{\Mt}}J_1(\W,\Mt) \ .
\end{eqnarray}

\def\PL{ {\it Phys. Lett.} }
\def\NP{ {\it Nucl. Phys.} }
\def\PTP{ {\it Prog. Theor. Phys.} }
\def\PRL{ {\it Phys. Rev. Lett.} }
\def\PR{ {\it Phys. Rev.} }

\begin{table}[htb]
\setlength{\tabcolsep}{.65cm}
\renewcommand{\arraystretch}{1.6}
        \caption[dummy]{ The $\D N\gamma$ form factors: $\a = - 
        {2 g_{\pi N\D} M_N^2\over F_\pi^2}$ and
         $[\bf I]$ is short hand notation
for $[g_A I_N + \chi g_1 I_\D]$, with $I\in\{F,J,H\}$ }
        \protect\label{table2}
\begin{tabular}{ cccc }
        \hline
        Source
    &  $G_1$
    &  $G_2$
    &  $G_3$
      \\ \hline
     Loops                                &
 ${\alpha\over M_N}[{\bf F}]\equiv \tilde{g}_1(q^2)$
& $-4\alpha [{\bf J}]\equiv \tilde{g}_2(q^2)$
& $-4\alpha \delta [{\bf H}]
\equiv\tilde{g}_3(q^2) $ \\ \hline
    ${\cal  A}^{(2)}_{\Delta N}      $    &  $\frac{b_1}{2}$ & $ b_1 $
                 & $ 0 $ \\ \hline
    $\gamma_0 {\cal B}_N^{\dagger(1)}\gamma_0
            {1\over {\cal C}_N^{(0)} }
          {\cal D}_{N\Delta}^{(2)}$  &
        $ 0 $ & $ -b_1-2b_6 $ & $ 0 $ \\ \hline
    $ \gamma_0 {\cal B}_{\Delta N}^{(2)\dagger}
         \gamma_0 {1\over {\cal C}_\Delta^{(0)}}
    {\cal B}^{(1)}_{\Delta}$  &
                   $ 0 $ & 0 & 0   \\ \hline
    ${\cal A}^{(3)}_{D_1}$     &
    ${\delta \over 4 M_N} D_1$ &  $ D_1$ &
    $ 0 $ \\ \hline
    ${\cal A}^{(3)}_{E_1}$  & $-{\delta \over 2 M_N} E_1$ & $ 0 $ &
    $ 0 $ \\ \hline
\end{tabular}\label{tab1}
\end{table}

\begin{table}[htb]
\setlength{\tabcolsep}{.35cm}
\renewcommand{\arraystretch}{1.6}
        \caption[dummy]{ The slope parameters [fm$^2$] of the form factors
$G_i(q^2)$}
        \protect
\begin{tabular}{ ccc }
        \hline
       $\rho^2_1$
    &  $\rho^2_2$
    &  $\rho^2_3$
      \\ \hline
                   $1.94 + 0.55\ i$
                 & $10.50 - 6.16\ i$
                 & $-4.12 - 4.48\ i$

      \\ \hline
\end{tabular}\label{tab2}
\end{table}


\begin{figure}[htb]
\begin{center}
\mbox{\epsfig{file=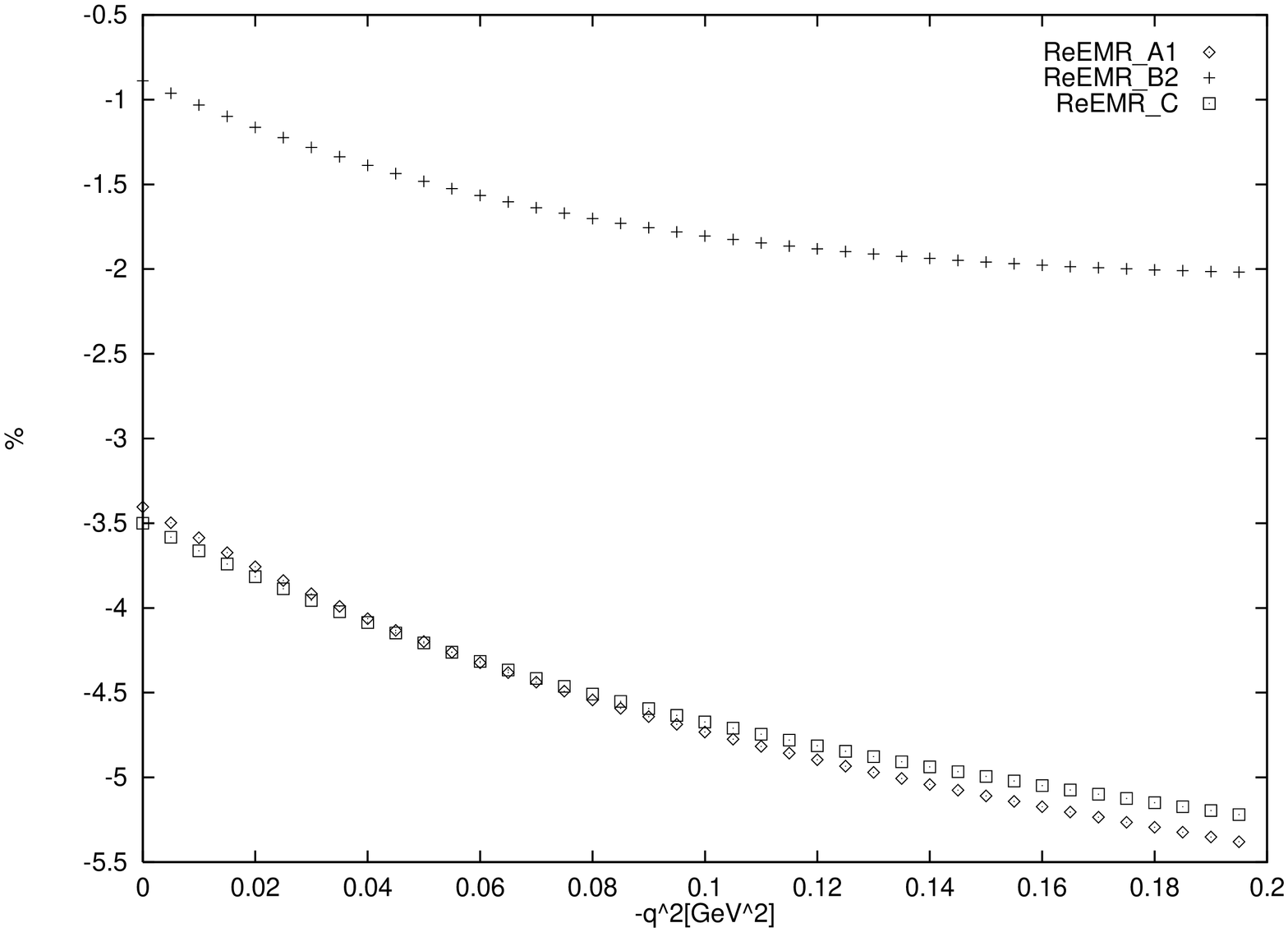,width=0.8\textwidth,angle=-90}}
\caption[dummy]{The real part of the EMR ratio
with $C_2$, $C_3$ given by the sets $A_1$, $B_2$, $C$.}
\label{fig9}
\end{center}
\end{figure}

\begin{figure}[htb]
\begin{center}
\mbox{\epsfig{file=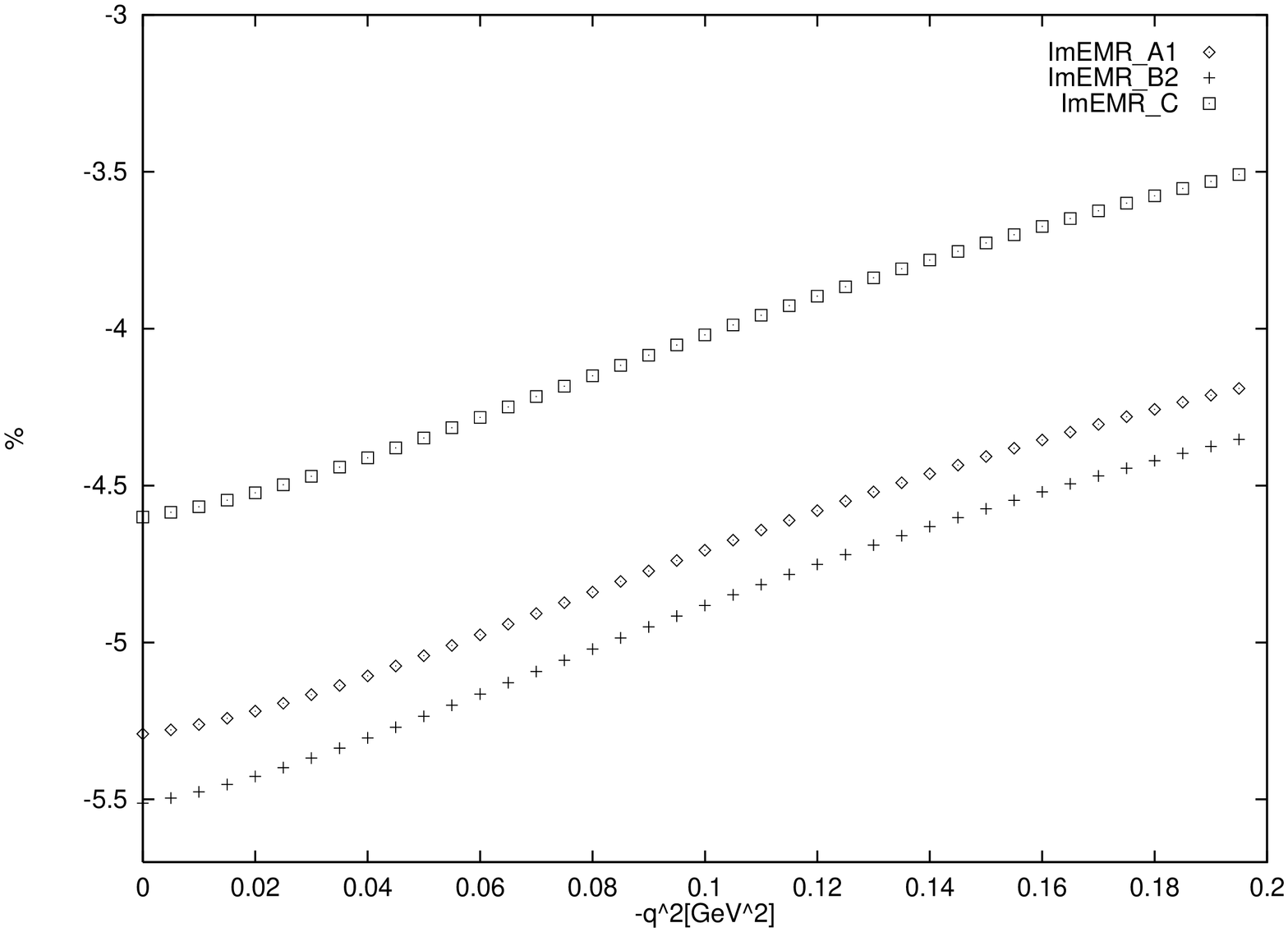,width=0.8\textwidth,angle=-90}}
\caption[dummy]{The imaginary part of the EMR ratio
with $C_2$, $C_3$ given by the sets $A_1$, $B_2$, $C$.}
\label{fig10}
\end{center}
\end{figure}

\begin{figure}[htb]
\begin{center}
\mbox{\epsfig{file=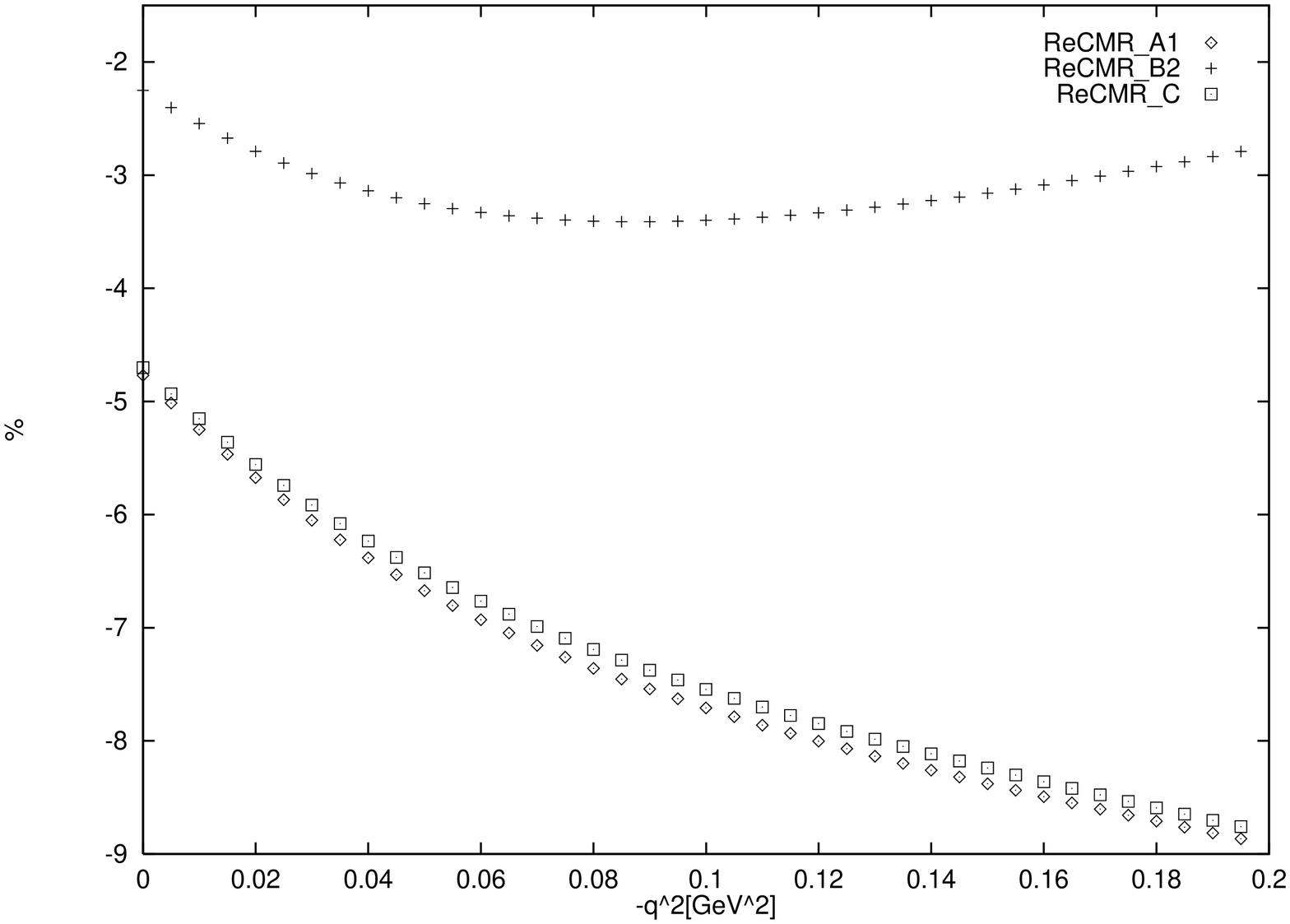,width=0.8\textwidth,angle=-90}}
\caption[dummy]{The real part of the CMR ratio
with $C_2$, $C_3$ given by the sets $A_1$, $B_2$, $C$.}
\label{fi11}
\end{center}
\end{figure}

\begin{figure}[htb]
\begin{center}
\mbox{\epsfig{file=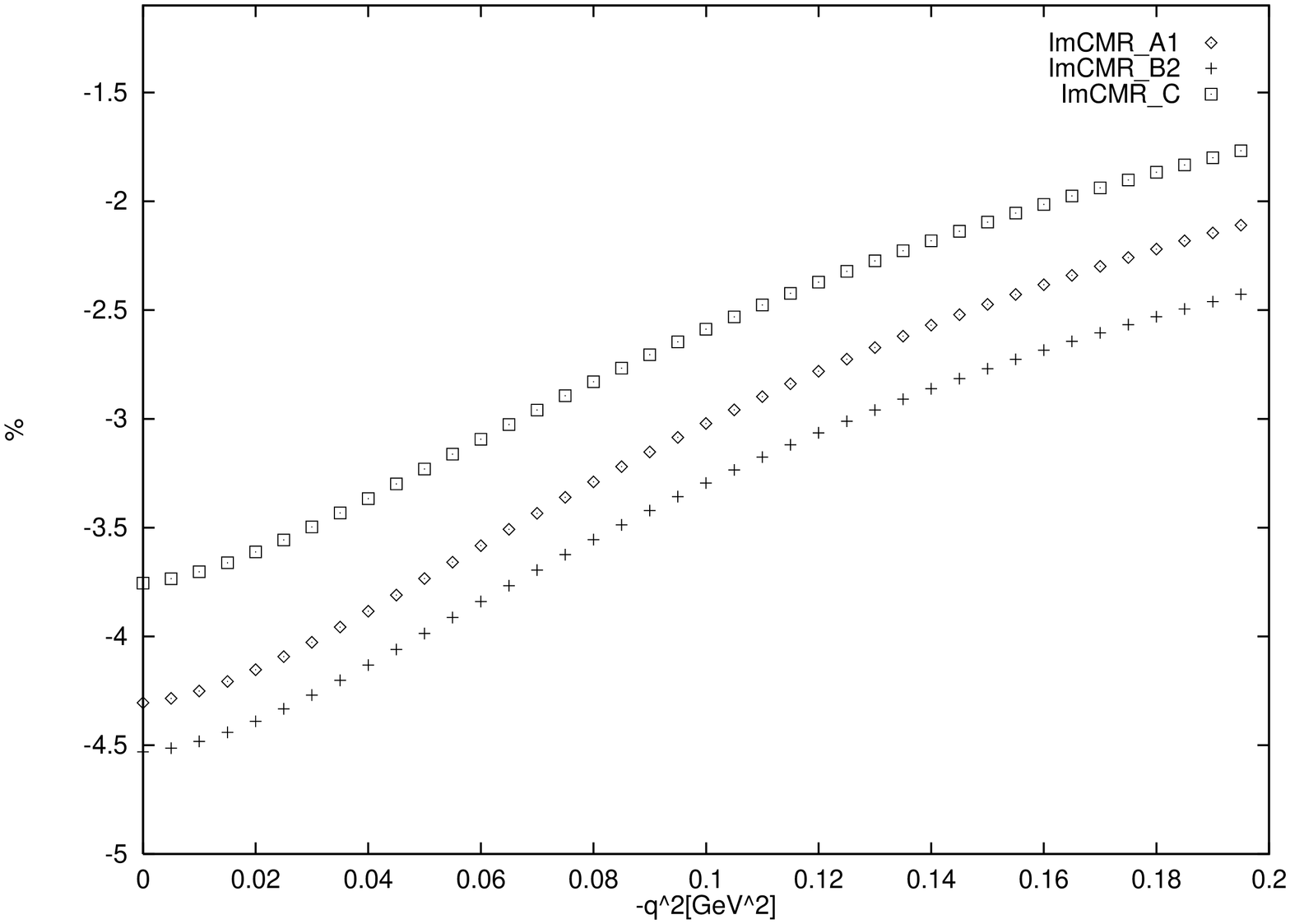,width=0.8\textwidth,angle=-90}}
\caption[dummy]{The imaginary part of the CMR ratio
with $C_2$, $C_3$ given by the sets $A_1$, $B_2$, $C$.}
\label{fig12}
\end{center}
\end{figure}


\begin{thebibliography}{99}

\bibitem{exp} R. Beck et al., Phys. Rev. Lett. 78, 606 (1997);
G. Blanpied et al., Phys. Rev. Lett. 79, 4337 (1997); F. Kalleicher et
al., Z. Phys. A359, 201 (1997).

\bibitem{th} R.M. Davidson, N.C. Mukhopadhyay and R.S. Wittman, Phys.
Rev. D43, 71 (1991); R.A. Arndt, I.I. Strakovsky and R.L. Workman, Phys. Rev.
C56, 577 (1997).

\bibitem{disp} O. Hanstein, D. Drechsel and L. Tiator, Phys. Lett.
B385, 45 (1996); Nucl. Phys. A632, 561 (1998).

\bibitem{currexp} P. Bartsch et al., ``C2/M1 in $p\rightarrow\Delta^+$ via
$p(\vec{e},e'\vec{p})\pi^0$, Talk given at Baryons 98, to be published in
the proceedings;
M.O. Distler et al., ``Recent Measurements of the
$\gamma^\ast p\rightarrow\Delta$ Response Functions at Bates'', {\it
ibid.};
R.W. Gothe et al., ``The $N\rightarrow\Delta$ Transition at Low Momentum
Transfer'', {\it ibid.}.

\bibitem{NRQM} C.M.~Becchi and G.~Morpurgo, Phys. Lett. 17, 352 (1965);
N.~Isgur and G.~Karl, Phys. Rev. D18, 4187 (1978);
N.\ Isgur, et al.,    Phys. Rev. D25, 2394 (1982).

\bibitem{trento} {\it e.g. see} Proceedings of ``Joint ECT*/Jefferson Lab
Workshop
on N* Physics and Non-perturbative QCD'', Eds. V. Burkert et al., to be
published
in {\em Few Body Physics}.

\bibitem{pap}  Bates-OOPS Collaboration, C.N. Papanicolas, spokesperson;
MAMI-A1 Collaboration, R. Neuhausen, spokesperson;
ELSA-ELAN Collaboration, B. Schoch, spokesperson;
BNL-LEGS Collaboration, A. Sandorfi, spokesperson.

\bibitem{cm} C.M. Carlson and N.C. Mukhopadhyay, Phys. Rev. Lett. 81, 2646
(1998).

\bibitem{XBM} G.~K\"albermann and J.M.~Eisenberg, Phys. Rev. D28, 71 (1984);
              K.~Bermuth et al, Phys. Rev. D37 (1988);
              D.~Lu et al., Phys. Rev. C55, 3108 (1997).

\bibitem{Buch} A. Buchmann, ``A Quark Model Calculation of the E1+/M1+
Ratio'', Talk given at Baryons 98, to be published in the proceedings.

\bibitem{Derek} D.B. Leinweber, et al.,  Phys. Rev. D48, 223 (1993).

\bibitem{Butler} M.N. Butler, M.J. Savage and R.P. Springer, Phys.
Lett. B304 353, (1993).

\bibitem{Lucio}M.\ Napsuciale and J.L. Lucio,
   Nucl. Phys. B494, 260 (1997).

\bibitem{Work} R.L. Workman, ``The E1+/M1+ Ratio, A Status Report'', Talk
given at Baryons 98, to be published in the proceedings.

\bibitem{short} T.R.\ Hemmert, B.R. Holstein and J. Kambor, Phys. Lett. B395,
89 (1997).

\bibitem{long} T.R.\ Hemmert, B.R. Holstein and J. Kambor , J. Phys. G24,
1831 (1998).

\bibitem{BKM} {\it e.g.} V. Bernard, N. Kaiser and U.-G. Mei{\ss}ner, Int. J.
of Mod.
Phys. E4, 193 (1995).

\bibitem{talk} T.R. Hemmert, ``$NN$ and $N\Delta$ Form Factors viewed from
ChPT'',
preprint no. nucl-th/9807050, to be published in \cite{trento}.

\bibitem{BFHM} V. Bernard, H.W. Fearing, T.R. Hemmert and U.-G. Mei{\ss}ner,
Nucl. Phys. A635, 121 (1998).

\bibitem{definition} H.F.~Jones and M.D.~Scadron, Ann. Phys. 81, 1 (1973);
   R.C.E. Devensish, T.S. Eisenschitz and J.G. K\"orner, Phys. Rev. D14,
3063 (1976); R.\ Davidson (private communication).

\bibitem{photo} V. Bernard, T.R. Hemmert and U.-G. Mei{\ss}ner,
``The Role of $\Delta$(1232) in Pion Photoproduction.'', forthcoming.

\bibitem{Weinb} S. Weinberg, Physica A96, 327 (1979).

\bibitem{GLeut} J. Gasser and H. Leutwyler, Ann. Phys. 158, 142 (1984).
              Nucl. Phys. B250, 465 (1985).

\bibitem{WIsgur} N. Isgur and M. Wise, Phys. Lett. B232, 113 (1989);
                 B237, 527 (1990).

\bibitem{Georgi} H. Georgi,  Phys. Lett. B240, 447 (1990).

\bibitem{JenkManoh} E. Jenkins and A. Manohar, Phys. Lett. B255, 558 (1991).

\bibitem{BKKM} V.Bernard, J. Kambor, N. Kaiser and U.-G. Mei{\ss}ner, Nucl.
Phys. B388, 315 (1992).

\bibitem{Finke} {\em e.g.} M. Finkemeier, H. Georgi and M. McIrvin, Phys.
Rev. D55, 6933 (1997).

\bibitem{Ecker} G. Ecker, preprint no. hep-ph/9805500 (1998).

\bibitem{GSS} J. Gasser, M.E. Sainio and A. Svarc, Nucl. Phys. B307, 779
(1988).

\bibitem{EM96} G. Ecker and M. Moijzis, Phys. Lett. B365, 312 (1996).

\bibitem{FMS} N. Fettes, U.-G. Mei{\ss}ner and S. Steininger, Nucl. Phys. A640,
199
(1998).

\bibitem{priv} U.-G. Mei{\ss}ner, G. M{\" u}ller and S. Steininger, preprint
no.
hep-ph/9809446 (1998).

\bibitem{ChPT97} ChPT97, Proc. of the Workshop on Chiral Dynamics --
Theory and Experiment, Mainz, Germany, Sept. 1-5, 1997;
A. Bernstein, D. Drechsel and T. Walcher (eds.), Springer Verlag,
Heidelberg.

\bibitem{JM2} E. Jenkins and A. Manohar, Phys. Lett. B259, 353 (1991).

\bibitem{kambor} J. Kambor, ``The Delta as an Effective Degree of Freedom in
Chiral Perturbation Theory''; in \cite{ChPT97}.

\bibitem{Compton2} T.R. Hemmert, B.R. Holstein, J. Kambor and G.
Kn{\" o}chlein, Phys. Rev. D57, 5746 (1998); T.R. Hemmert, B.R. Holstein and
J. Kambor, Phys. Rev. D55, 5598 (1997).

\bibitem{EW} {\it e.g.} see T. Ericson and W. Weise, Pions and Nuclei,
Oxford Science Publications, Oxford (1998).

\bibitem{VPI} R.L. Workman and R.A. Arndt, ``Stability of the E2/M1 ratio at
the T-matrix
pole'', preprint no. nucl-th/9809049 (1998).

\bibitem{RPI} R.M. Davidson et al., preprint no. nucl-th/9810038 (1998).

\bibitem{couplings} T.R. Hemmert, B.R. Holstein and N.C. Mukhopadhyay, Phys.
Rev. D51,
158 (1995).








\end{thebibliography}
\end{document}